\documentclass{emulateapj}
\usepackage{color} 

\usepackage{epsfig}
\usepackage{subfigure}
\usepackage{graphicx}
\usepackage{amsmath}
\usepackage{natbib}
\newcommand{\pa}{\partial}
\newcommand{\mb}{\boldsymbol}
\newcommand{\mc}{\mathcal}

\newcommand{\actaa}{Acta Astronomica}

\shorttitle{Stratified MRI in Strong Vertical Field}
\shortauthors{X.-N. Bai \& J. M. Stone}

\begin{document}


\title{Local Study of Accretion Disks with a Strong Vertical Magnetic Field:
Magnetorotational Instability and Disk Outflow}


\author{Xue-Ning Bai\altaffilmark{1,2,3} \& James M. Stone\altaffilmark{1}}

\altaffiltext{1}{Department of Astrophysical Sciences, Peyton Hall, Princeton
University, Princeton, NJ 08544}
\altaffiltext{2}{Hubble Fellow}
\altaffiltext{3}{Current address: Institute for Theory and Computation,
Harvard-Smithsonian Center for Astrophysics, 60 Garden St., MS-51, Cambridge, MA 02138}

\email{xbai@cfa.harvard.edu}




\begin{abstract}
We perform three-dimensional vertically-stratified local shearing-box ideal MHD
simulations of the magnetorotational instability (MRI) that include a net vertical
magnetic flux, which is characterized by midplane plasma $\beta_0$ (ratio of gas to
magnetic pressure). We have considered $\beta_0=10^2, 10^3$ and $10^4$ and in
the first two cases the most unstable linear MRI modes are well resolved in the
simulations. We find that the behavior of the MRI turbulence strongly depends on
$\beta_0$: The radial transport of angular momentum increases with net vertical flux,
achieving $\alpha\sim0.08$ for $\beta=10^4$ and $\alpha\gtrsim1.0$ for
$\beta_0=100$, where $\alpha$ is the height-integrated and mass-weighted
Shakura-Sunyaev parameter. A critical value lies at $\beta_0\sim10^3$:
For $\beta_0\gtrsim10^3$, the disk consists of a gas pressure dominated midplane
and a magnetically dominated corona. The turbulent strength increases with net flux,
and angular momentum transport is dominated by turbulent fluctuations. The
magnetic dynamo that leads to cyclic flips of large-scale fields still exists, but
becomes more sporadic as net flux increases. For $\beta_0\lesssim10^3$, the
entire disk becomes magnetic dominated. The turbulent strength saturates, and the
magnetic dynamo is fully quenched. Stronger large-scale
fields are generated with increasing net flux, which dominates angular momentum
transport. A strong outflow is launched from the disk by the magnetocentrifugal
mechanism, and the mass flux increases linearly with net vertical flux and shows sign
of saturation at $\beta_0\lesssim10^2$. However, the outflow is unlikely to be {\it directly}
connected to a global wind: for $\beta_0\gtrsim10^3$, the large-scale field has no
permanent bending direction due to dynamo activities, while for $\beta_0\lesssim10^3$,
the outflows from the top and bottom sides of the disk bend towards opposite directions,
inconsistent with a physical disk wind geometry. Global simulations are needed to
address the fate of the outflow.
\end{abstract}



\section{Introduction}\label{sec:intro}

The magnetorotational instability (MRI, \citealp{BH91}) is believed to be the
primary mechanism for driving accretion in a wide range of astrophysical disks
by generating turbulence to provide outward angular momentum transport. The
robustness of the MRI and its physical properties have been widely studied by
means of both local shearing box simulations (e.g., \citealp{HGB95,Brandenburg_etal95})
and global simulations (e.g., \citealp{Hawley01,FromangNelson06}). Local shearing-box
simulations have the advantages of being able to achieve high resolution at
modest computational cost, and are the primary way for exploring the detailed
physics of the MRI.

Four types of shearing-box MRI simulations exist, depending on whether vertical
gravity from the central object (which gives density stratification) is included and
whether the simulations include net vertical magnetic flux. While unstratified MRI
simulations provide important benchmarks on the physical properties of the MRI
turbulence (e.g.,
\citealp{HGB96,Sano_etal04,FromangPap07b,Fromang_etal07, LesurLongaretti07}),
real disks involve vertical density stratification. So far most vertically stratified
shearing-box simulations focus on a magnetic field configuration with zero net vertical
magnetic flux (e.g., \citealp{SHGB96,MillerStone00,ZieglerRudiger00,Hirose_etal06,
Davis_etal10,Shi_etal10,Flaig_etal10,Simon_etal12a}). However, the inclusion
of net vertical magnetic field in stratified shearing-box simulations, which is
probably closer to reality, has only been rarely explored, most likely due to
numerical difficulties \citep{SHGB96,MillerStone00}: either the numerical code
crashes or the disk is violently disrupted in a few orbits.

More and more attention has been drawn to this least explored regime of stratified
shearing-box with net vertical flux in the recent years. This is partly due to the
improvement of the numerical schemes, but more importantly, the physical
significance of the MRI in this regime and its intimate connections to astrophysical
applications make it deserve more detailed study. In particular, the presence of
jets and outflows in a wide range of accretion systems such as protostars
(e.g., \citealp{ReipurthBally01,Cabrit07}),
X-ray binaries (e.g., \citealp{MirabelPodriguez99,Fender06})
and active galactic nuclei (e.g., \citealp{Begelman_etal84,HarrisKrawczynski06})
is indicative of the presence of large-scale (poloidal) magnetic field
threading through the disk, which is essential for jet/outflow acceleration and
collimation via the magnetocentrifugal mechanism
\citep{BlandfordPayne82,PudritzNorman83}.

Recently, \citet{SuzukiInutsuka09} and \citet{Suzuki_etal10} successfully
conducted stratified shearing-box simulations that include net vertical magnetic
field with an outflow boundary condition using characteristic decomposition. Their
net vertical field, characterized by $\beta_0$ (ratio of gas to magnetic pressure at
midplane), is rather weak ($\beta_0\geq10^4$). They reported the launching of
disk outflow and found that the averaged mass outflow rate increases linearly with
the magnetic pressure of the net vertical field. It was speculated that the outflow
serves as the base of a Blandford-Payne type magneto-centrifugal wind.

In this paper, we conduct shearing-box simulations of the MRI that include a
relatively strong net vertical magnetic flux, with $10^4\geq\beta_0\geq100$
which has largely been unexplored in the literature.
The main motivation of our work is two-fold as we elaborate below.

First, we aim at studying the properties of the MRI turbulence in more realistic
magnetic field geometry and their dependence on the net poloidal magnetic flux
threading through accretion disks. In particular, how the Shakura-Sunyaev
$\alpha$ parameter, which is of the most common interest, depends on the net
magnetic flux. Due to the vast dominance of zero net vertical flux stratified
shearing-box simulations, the $\alpha$ parameter has been taken for granted to
be of the order $0.01$, while the $\alpha$ value inferred from observations of
fully ionized disks is at least an order of magnitude higher (e.g., see discussions
by \citealp{King_etal07} and references therein). Unstratifed shearing-box
simulations (e.g., \citealp{HGB95,BaiStone11}) unambiguously found that as
long as the most unstable linear MRI mode fits into the simulation domain, the
turbulent stress increases roughly linearly with $1/\beta_0$. It is natural to
expect the same trend in stratified shearing-box simulations, hence net vertical
magnetic flux is likely to be a crucial ingredient in real accretion disks.
Moreover, as in the zero net-flux cases, we expect buoyancy and open vertical
boundaries in our stratified simulations to give rise to novel features on the
properties of the MRI turbulence, particularly the MRI dynamo and the generation
of large-scale fields \citep{LesurOgilvie08,Gressel10,GuanGammie11}.

Second, our study will address the potential connection between the MRI and
the magneto-centrifugal wind (MCW, \citealp{BlandfordPayne82}). The MCW
extracts both mass and angular momentum from accretion disks, and is a
competing mechanism for driving disk accretion and evolution. While the wind
acceleration and collimation of the MCW is largely known, the wind launching
process which loads mass from the disk onto the wind is still not well understood.
Most semi-analytical works and numerical simulations either assume a razor-thin
disk treated as a boundary condition with artificial mass injection
(e.g., \citealp{Li95,OuyedPudritz97a,Krasnopolsky_etal99}),
or proceed with unresolved disk by adopting some forms of artificial diffusion,
(e.g., \citealp{Kato_etal02,CasseKeppens02,Zanni_etal07}).
Such diffusion, which presumably arises from the MRI, is necessary to allow
the gas to slide through the disk that avoids rapid accumulation of magnetic
flux to maintain steady state accretion. However, the microphysics (e.g., the
MRI) within the accretion disk, which is crucial for wind launching, is not modeled
directly. In fact, it is numerically very challenging to
properly resolve the MRI turbulence in global wind simulations, which
requires extended three-dimensional computational domain with large
numerical resolution. On the other hand, local shearing-box simulations
provide superb resolution with realistic computational cost, hence are a
powerful tool for studying the wind launching process from first principle.


Recently, \citet{Fromang_etal12} conducted a series of stratified shearing-box
MRI simulations with $\beta_0=10^4$. After carefully examining various numerical
issues, they reported simultaneous MRI turbulence and launching of an MCW,
whereas angular momentum transport is still dominated by the MRI rather than the
MCW. In parallel, \citet{Lesur_etal12} conducted a series of stratified shearing-box
simulations in both one-dimension and three-dimensions with $\beta_0\sim10$,
where the most unstable linear MRI modes just marginally fit into the disk. They
found that such MRI modes directly produce magnetically driven outflows, which
are time varying due to secondary instabilities. Meanwhile, \citet{Moll12} studied
the wind launching process in two-dimensional shearing-box simulations with
$\beta_0\sim1-10$ (where MRI is suppressed) and found long-wavelength
``clump" instability for $\beta_0\lesssim1$ and speculated its connection to the
mass-flux instability found in earlier analytical works
\citep{Lubow_etal94,CaoSpruit02}. Our simulations fill the gap in parameter
space explored by these authors and reveal the interesting transition from the
MRI-dominated transport to potentially wind-dominated transport. Together with
the earlier semi-analytical study of wind launching by \citet{Ogilvie12}, which
applies to the regime of $\beta_0\lesssim1$, as well as the simulations by
\citet{SuzukiInutsuka09}, which applies to $\beta_0\gtrsim10^4$, the series
of studies cover the complete parameter range relevant to MRI turbulence and
wind launching in realistic accretion disks under the local shearing-box framework.

This paper is organized as follows. In Section \ref{sec:method} we describe
the methodology and setup of our simulations. We present our simulation results
in Section \ref{sec:results}, focusing on the properties of the MRI and angular
momentum transport, and Section \ref{sec:wind}, focusing on the dynamics of
the outflow and its connection to global disk winds. Each result section is
broken into three subsections, devoting to one aspect of the results.
Implications of our results for global disk evolution are discussed in Section
\ref{sec:conclusion} before we conclude.

\section[]{Simulation Setup}\label{sec:method}

We use the Athena magnetohydrodynamic (MHD) code \citep{Stone_etal08} and
perform three-dimensional (3D) local MHD simulations of gas dynamics using the
shearing-box approach \citep{GoldreichLyndenBell65}. Ideal MHD equations are
written in a Cartesian coordinate system in the corotating frame at a fiducial radius
$R$ with Keplerian frequency $\Omega$, with ${\mb e}_x,{\mb e}_y,{\mb e}_z$
denoting unit vectors pointing to the radial, azimuthal and vertical directions
respectively, and ${\mb\Omega}$ is along the ${\mb e}_z$ direction. The MHD
equations read
\begin{equation}\label{eq:gascont}
\frac{\pa\rho}{\pa t}+\nabla\cdot(\rho\mb{u})=0\ ,
\end{equation}
\begin{equation}
\frac{\pa\rho\mb{u}}{\pa t}+\nabla\cdot(\rho\mb{u}^T{\mb u}
+{\sf T})=\rho\bigg[2{\mb u}\times{\mb\Omega}
+3\Omega^2x{\mb e}_x-\Omega^2z{\mb e}_z\bigg]\ ,
\label{eq:gasmotion}
\end{equation}
\begin{equation}
\frac{\pa{\mb B}}{\pa t}=\nabla\times({\mb u}\times{\mb B})\ ,\label{eq:induction}
\end{equation}
where ${\sf T}$ is the total stress tensor
\begin{equation}
{\sf T}=(P+B^2/2)\ {\sf I}-{\mb B}^T{\mb B}\ ,
\end{equation}
$\rho$ is the gas density, ${\mb u}$ is the gas velocity. We use an isothermal
equation of state $P=\rho c_s^2$. Note that the unit for magnetic field is such that
the magnetic pressure equals $B^2/2$, which avoids the extra factor of
$\sqrt{4\pi}$. The vertical gravity from the central object ($-\Omega^2z$) is included
in the momentum equation hence our simulations have vertical density stratification.

The shearing-box source terms (Coriolis force and tidal gravity) have been readily
implemented in Athena \citep{StoneGardiner10}. An orbital advection scheme is
adopted where the system is split into an advective part that corresponds to the
background shear flow $-3\Omega x/2{\mb e}_y$, and the other part that only
evolve the velocity fluctuations ${\mb v}$:
\begin{equation}
{\mb v}\equiv{\mb u}+\frac{3}{2}\Omega x{\mb e}_y\ .
\end{equation}
Most of our analysis deals with ${\mb v}$, except in the study of conservation
laws along field lines, where ${\mb u}$ will also be used (see Section
\ref{sssec:conslaw}).

We use standard shearing-box boundary conditions in the radial direction
\citep{HGB95}. In the vertical direction, we adopt a simple zero-gradient outflow
boundary condition for velocities and magnetic fields, with density extrapolated
assuming vertical hydrostatic equilibrium \citep{Simon_etal11}. In addition,
vertical velocity in the ghost cells is set to zero in the case of inflow. It was
pointed out that strict zero-gradient boundary condition (including zero vertical
density gradient) would prevent MHD-driven outflow \citep{Lesur_etal12}, while
we find that the outflow can be launched once a vertical density gradient is
present at the interface of the vertical boundary.

It is well known that in the presence of pure vertical background magnetic field, the
linear MRI modes consist of counter-motions in the radial direction that alternate
along the vertical direction (i.e., the channel flow). The channel flow is an exact
solution of the MHD equations even in the non-linear regime provided that the
magnetic field remains sub-thermal \citep{GoodmanXu94}.
It can achieve very large amplitude before broken up by parasitic instabilities
\citep{Latter_etal09,PessahGoodman09}. Recently, \citet{Latter_etal10} studied
the MRI modes in the presence of vertical stratification and their parasitic
instabilities. They found that the channel modes strongly amplify the magnetic
fields to thermal strength before they can be destroyed. On the simulation side, the
channel flows in the initial stage of the simulations not only place enormous demands
on numerical codes, but also give rise to eruptive behavior that depletes the disk in
just a few orbits \citep{MillerStone00}. In view of the transient nature of the channel
flow, as well as its large numerical demand, we initialize our simulations in a way
that strong channel flow can be avoided, as described below.


Fiducially, we use a box size of ($4H, 8H, 12H$) in ($x, y, z$) resolved by
($96, 96, 288$) grid points, where $H\equiv c_s/\Omega$ is the thermal
scale height. The relatively large horizontal box size is necessary to properly
capture the mesoscale structures of the MRI turbulence such as the zonal flow
\citep{Johansen_etal09a,Simon_etal12a}. The initial density profile is taken to be
Gaussian $\rho=\rho_0\exp{(-z^2/2H^2)}$. We use natural unit in our
simulations, with $c_s=1$, $\Omega=1$, $H=1$ and $\rho_0=1$. We
initialize our simulations with zero net vertical magnetic flux configuration, i.e., a
weak vertical magnetic field that varies sinusoidally in the radial ($x$) direction.
We run the zero net-flux simulation to $t=120\Omega^{-1}$ when turbulence is
fully developed, then we start to add net vertical magnetic flux, which is applied
uniformly in the grid (hence it does not introduce any divergence error) and the
amount added is proportional to the time step d$t$. This process continues to
$t=240\Omega^{-1}$ when the net vertical magnetic field strength reaches the
desired level $B_0$, characterized by midplane plasma
$\beta_0=2\rho_0c_s^2/B_0^2$. The simulations are run for about
another 150 orbits to $t=1200\Omega^{-1}$.

For our simulations, we consider $\beta_0=10^2, 10^3$ and $10^4$, corresponding
to runs B2, B3 and B4. Ignoring vertical stratification, the most most unstable MRI
wavelength at disk midplane is about $9.18\beta_0^{-1/2}H$ \citep{HGB95}. Including
vertical stratification, calculations by \citet{Latter_etal10} suggest resolution of
about 25 cells per $H$ for $\beta_0=100$, and 50 cells per $H$ for $\beta=10^3$, to
properly resolve the MRI channel modes. Our fiducial resolution meets this criterion
for $\beta_0=100$. For $\beta_0=10^3$, we conduct an additional simulation (run
B3-hr) with twice the resolution which then does properly resolve the most
unstable linear modes. For this run, we initialize the simulation with full a net vertical
magnetic flux, together with a sinusoidally varying vertical field component so as to
avoid an excessively strong initial channel flow. The list of simulation runs is provided
in Table \ref{tab:runs}.

\begin{table}
\caption{Summary of Simulation Runs.}\label{tab:runs}
\begin{center}
\begin{tabular}{ccccc}\hline\hline
 Run & $\beta_0$ & Domain Size & Resolution & Duration$^1$ \\\hline
B2 & $10^2$ & $4H\times8H\times12H$ & $96\times96\times288$ & $960\Omega^{-1}$ \\
B3 & $10^3$ &  $4H\times8H\times12H$ & $96\times96\times288$ & $960\Omega^{-1}$ \\
B4 & $10^4$ &  $4H\times8H\times12H$ & $96\times96\times288$ & $960\Omega^{-1}$ \\
B3-hr & $10^3$ &  $3H\times6H\times12H$ & $144\times144\times576$ & $720\Omega^{-1}$ \\
\hline\hline
\end{tabular}
\end{center}
$^1$: duration of the run with net flux fully added.
\end{table}

A density floor of $\rho_{\rm Floor}=10^{-4}, 3\times10^{-5}$ and $10^{-5}$ are
applied for simulations with $\beta_0=10^2, 10^3$ and $10^4$ respectively to avoid
numerical difficulties at magnetically dominated (low plasma $\beta$) regions. We will
see in Figure \ref{fig:profile} that horizontally averaged densities in the saturated
states of all our simulations are at least two orders of magnitude larger than the
density floor. In addition, we apply another density floor that is ten times smaller than
$\rho_{\rm floor}$ to all the left and right states inside the MHD integrator (we use the
CTU integrator with third order spatial reconstruction, see \citealp{Stone_etal08} for
details). This procedure is essential and greatly improves the robustness of the
Athena MHD code.

\begin{figure}
    \centering
    \includegraphics[width=105mm]{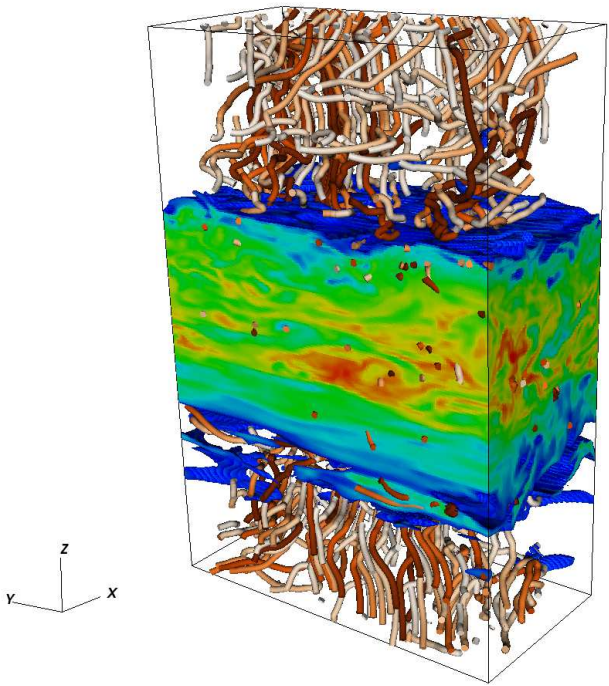}
  \caption{One snapshot from our simulation run B3. The surface plot shows the
  isosurface of gas density, increasing logarithmically from blue with $\rho=0.1$
  (cutoff) to red with $\rho=2.0$. The streamlines indicate gas velocity, increasing
  linearly from white with $v=0$ to dark red with $v=5c_s$.}\label{fig:snapshot}
\end{figure}

Our simulations with $\beta_0=10^2$ and $10^3$ occasionally produce extremely
strong magnetic field in very localized regions (a few grid cells in the entire
computational domain), which severely limits the simulation time step. In order to
proceed, we have included some hyper-resistivity $\eta\propto J^2$ in regions with
excessively large $J$, where $J$ is the current density. More specifically, we define
$j_d\equiv|\nabla\times{\mb B}|/\sqrt{\rho}$, and set Ohmic resistivity
$\eta=0.1(c_s^2/\Omega)\cdot(j_d/10^3\Omega)^2$ when $j_d>10^2\Omega$ and
$\eta=0$ otherwise. We have examined that for the $\beta_0=10^2$ and $10^3$
cases, hyper-resistivity is applied to at most $1\%$ of all grid cells (most of the time
less than $0.1\%$) and does not affect the general properties of the MRI turbulence.

Our simulations always launch a strong outflow (see Figure \ref{fig:wind}) that would
drain the gas in the simulation box in tens to hundreds of orbits. In order for the
system reach a steady state which allows us to characterize the dynamical structure
of the turbulent disk, we enforce the total mass contained in the simulation domain to
be constant by multiplying the density of all grid cells by a common factor after each
time step. The momentum remains fixed in each grid cell. This approach is similar to
\citet{Ogilvie12} who considered much stronger magnetic fields that would launch
even stronger disk outflow. As will be discussed briefly in Section \ref{sssec:energetics},
such mass addition does not significantly affect the energetics of the system for all
our simulation runs.

In Figure \ref{fig:snapshot} we show a snapshot of our simulation run B3 in the
saturated state of the MRI turbulence. We see clearly the large density fluctuations
as a result of the vigorous MRI turbulence, and a substantial fraction of the gas
is in supersonic motion. Meanwhile, there is a systematic vertical velocity at the
vertical boundaries, indicating a strong outflow. In the following two sections, we
discuss in detail the properties of the MRI turbulence and the outflow respectively.

\section[]{Simulation Results: MRI Turbulence}\label{sec:results}

One of the main diagnostics for the MRI turbulence is the ${r\phi}$ component of
the stress tensor, consisting of the sum of Reynolds and Maxwell stresses,
\begin{equation}
T_{r\phi}^{\rm Rey}=\rho v_xv_y\ ,
\qquad T_{r\phi}^{\rm Max}=-B_xB_y\ ,\label{eq:alpha1}
\end{equation}
and they characterize the rate of radial angular momentum transport. We are
interested in the horizontally averaged vertical profiles (notated with an over bar,
here $\overline{T}_{r\phi}$) of the stresses, and they are usually normalized by
$\rho_0c_s^2$. Another useful measure is the
mass weighted average of the stresses, normalized to the gas pressure
\citep{ShakuraSunyaev73}
\begin{equation}
\alpha=\frac{\int\overline{T}_{r\phi}(z)dz}{\int\bar{\rho}(z)c_s^2dz}\ ,\label{eq:alpha2}
\end{equation}
where $\alpha$ denotes either Reynolds or Maxwell stresses (with subscript), or
their sum (without subscript), which characterizes the total rate of radial angular
momentum transport in the disk.

\begin{figure}
    \centering
    \includegraphics[width=85mm]{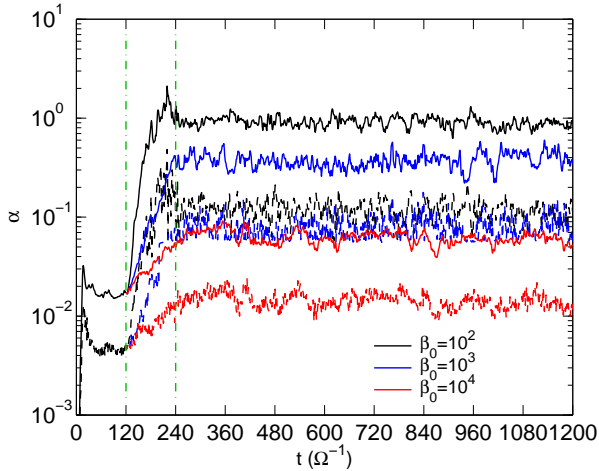}
  \caption{Time history of mass weighted Maxwell (solid) and Reynolds (dashed)
  stresses from simulations B2, B3 and B4. They are restarted from the end of the
  zero net-flux simulation at $t=120\Omega^{-1}$. Net vertical magnetic field is
  added gradually to its full strength at $t=240\Omega^{-1}$.}\label{fig:history}
\end{figure}

We show the evolution of $\alpha_{\rm Rey}$ and $\alpha_{\rm Max}$ in Figure
\ref{fig:history}. The total stress in the initial zero net-flux simulation saturates at
$\langle\alpha_{\rm Rey}\rangle\approx5.0\times10^{-3}$ and
$\langle\alpha_{\rm Max}\rangle\approx2.0\times10^{-2}$ respectively, consistent
with previous works (e.g., \citealp{Simon_etal12a}). The stresses increase steadily
as we gradually add net vertical magnetic flux from $t=120\Omega^{-1}$, and saturate
as full net flux is in place at $t=240\Omega^{-1}$. We wait for about another $20$
orbits and take the data from $t=360\Omega^{-1}$ to the end of the simulations
at $t=1200\Omega^{-1}$ for all the time averages in this section, which amounts to
about $130$ orbits. For our high-resolution run B3-hr, we consider data from
$t=180\Omega^{-1}$ to the end of the simulation at $t=720\Omega^{-1}$, which
amounts to about 85 orbits. Throughout this paper, we use angle bracket
$\langle\cdot\rangle$ to denote time average.

In the saturated state, a common indicator for proper resolution of the MRI
turbulence and convergence is the quality factor, defined as
\begin{equation}
Q_j\equiv\frac{2\pi|v_{Aj}|}{\Omega\Delta x_j},\ \quad{\rm where}\quad
|v_{Aj}|\equiv\sqrt{\overline{B_j^2}/\overline{\rho}}\ ,
\end{equation}
which is the ratio of a characteristic MRI wavelength to the grid scale. Here $j$
denotes either the $\hat{\mb y}$ or the $\hat{\mb z}$ dimension. \citet{Sano_etal04}
found that $Q_z\gtrsim6$ is required for the MRI turbulence to be well resolved.
\citet{Hawley_etal11} further suggested that $Q_z\gtrsim10$ and $Q_y\gtrsim20$
suffice for properly resolving the MRI turbulence. Although these results are mainly
based on either unstratified simulations or stratified simulations with zero net
vertical magnetic flux, we take them as useful reference. In our simulations, the
quality factors minimize at the disk midplane, where we find $\langle Q_z\rangle$
to be about $87$, $50$, $12$ for runs B2, B3 and B4 respectively, and
$\langle Q_y\rangle$ is about $185$, $110$, $25$ for the three runs. For the
high-resolution run B3-hr, we find $\langle Q_z\rangle\sim110$ and
$\langle Q_y\rangle\sim285$. In all cases, the quality factors are well above the
suggested value.

\subsection[]{Disk Structure}

\subsubsection[]{Vertical Profiles}

\begin{figure}
    \centering
    \includegraphics[width=85mm]{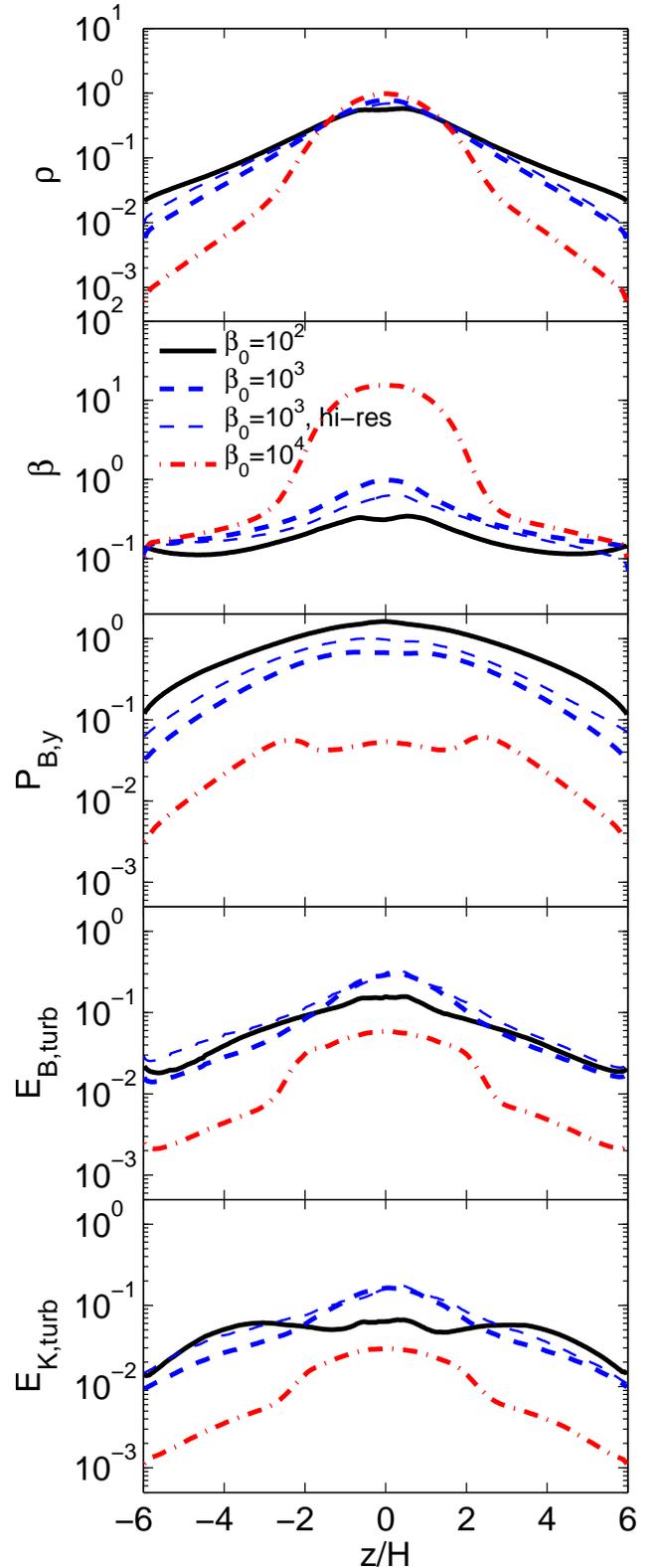}
  \caption{Vertical profiles of (from top to bottom) gas density $\langle\bar\rho\rangle$,
  plasma $\langle\bar\beta\rangle$, magnetic pressure of the toroidal field
  $\langle\bar{P}_{B, y}\rangle$,  turbulent magnetic energy
  $\langle\bar{E}_{\rm B,turb}\rangle$ and turbulent kinetic energy
  $\langle\bar{E}_{\rm K,turb}\rangle$. In each panel, we show the profiles for runs
  B2 ($\beta_0=10^2$, solid), B3, ($\beta_0=10^3$ dashed) and B4 ($\beta_0=10^4$,
  dash-dotted) respectively, and the thin dashed lines correspond to run B3-hr
  ($\beta_0=10^3$ at high resolution).}\label{fig:profile}
\end{figure}

In the presence of a strong net vertical magnetic flux, the systems all saturate into
very powerful MRI turbulence that substantially changes the structure of the disk. 
In Figure \ref{fig:profile} we show the vertical profiles of various horizontally averaged
quantities. The upper two panels are for gas density $\langle\bar{\rho}\rangle$ and
plasma $\langle\bar{\beta}\rangle$ (ratio of gas to magnetic pressure). We see that the
density profile deviates substantially from Gaussian due to strong magnetic support.
Only for $\beta_0=10^4$, the Gaussian kernel is still present; while for
$\beta_0\leq10^3$, the entire disk becomes magnetically dominated, with plasma
$\beta\lesssim1$ everywhere. In all cases, the averaged plasma $\beta$ saturates at
about 0.1 towards the disk surface, while in the extreme situation of $\beta_0=10^2$,
plasma $\beta$ is about $0.1$ everywhere as MRI saturates. The magnetic pressure
profile shows a large vertical gradient when $\beta\lesssim1$ (as can be seen by
dividing the first two panels of Figure \ref{fig:profile}). The magnetic pressure is
dominated by contributions from the toroidal field (as can be compared with the third
panel of the Figure), especially when $\beta_0\lesssim10^3$. Correspondingly, the
disk is substantially puffed up\footnote{We note that \citet{Lesur_etal12} and
\citet{Moll12} observed strong compression of the disk in their simulations where
stronger vertical field ($\beta_0\leq10$) is used and reflection symmetry about the
midplane is assumed. The compression is due to the large poloidal field curvature
around the disk midplane. We do not find any disk compression in the case of
$\beta_0\geq10^2$ mainly because no reflection symmetry is enforced to build up the
poloidal field curvature. In fact, our simulation results show opposite symmetry across
the midplane and the poloidal field curvature near midplane is about zero. See further
discussions about symmetry in Section \ref{ssec:fate}.}. We also note a potential
caveat: the fact that the entire disk becomes magnetically dominated implies that in
real disks, there can be strong magnetic pressure gradient support in the radial
direction. This may lead to substantial sub-Keplerian rotation and in turn modifies the
behavior of the MRI, a situation that requires global simulations to be properly
addressed (see further discussions in Section \ref{ssec:global}).

In the lower two panels of Figure \ref{fig:profile} we show the profiles of horizontally
averaged turbulent magnetic energy $\langle\bar{E}_{\rm B, turb}\rangle$ and kinetic
energy $\langle\bar{E}_{\rm K, turb}\rangle$. Here we have subtracted contributions
from horizontally averaged mean fields and mean velocities at each vertical layer in
the calculations: $\bar{E}_{\rm K, turb}\equiv\bar{E}_K-\bar{\rho}(\bar{v})^2/2$,
$\bar{E}_{\rm B, turb}\equiv\bar{E}_B-(\bar{B})^2/2$.
From the figure we see that for both magnetic and kinetic components, turbulent
energy increases substantially by about one order of magnitude as $\beta_0$ changes
from $10^4$ to $10^3$. As net vertical field increases further, to $\beta_0=10^2$,
however, turbulent energy roughly stays at the same level as the $\beta_0=10^3$ case.
This indicates that the strength of the MRI turbulence saturates as
$\beta_0\lesssim10^3$. On the other hand, the amplitude of the mean flow and mean
magnetic field continue to increase with net vertical magnetic field, and their
contributions to the total energy gradually dominate that from turbulent energy. We will
discuss this fact further in Section \ref{ssec:amt}. Here we note another caveat that
in the presence of strong mean toroidal field at the level of equipartition or higher,
the linear dispersion relation of the MRI is strongly modified by field curvature
\citep{PessahPsalties05}, which is ignored in the shearing-box approximation, and
again requires global simulations to be addressed properly.

\begin{figure}
    \centering
    \includegraphics[width=90mm]{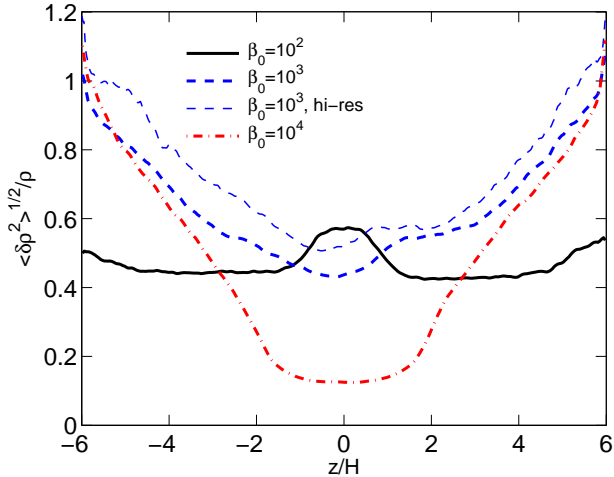}
  \caption{Vertical profiles of density fluctuations $\langle\delta\rho^2\rangle^{1/2}(z)$
  normalized to horizontally averaged density $\langle\bar\rho\rangle$ for all simulation
  runs.}\label{fig:drho}
\end{figure}

For $\beta_0\lesssim10^3$, turbulence in all regions of the disk become trans-sonic
or super-sonic, as one compares the profiles $\langle\bar{E}_{\rm K, turb}\rangle$ and
$\langle\bar\rho\rangle$.
This result implies strong compression and rarefaction, hence large density variations.
Looking at the density field, we find that the density fluctuations in our simulations generally
show elongated structure that is slightly tilted anti-clockwise from the azimuthal direction,
indicating spiral density waves excited by the MRI turbulence \citep{HeinemannPap09b},
although these structures are short lived due to the strong turbulence. 
To address such density variations, we measure the standard deviations of
density fluctuations $\delta\rho$ in each horizontal layer, and normalize them to the
horizontally averaged densities. The obtained profile of
$\langle\delta\rho^2\rangle^{1/2}/\rho$ for all simulations are shown in Figure
\ref{fig:drho}. We see that density variations are significant in all cases. For runs B3
and B4, the midplane density fluctuation increases with net magnetic flux, and reaches
about $0.5$ for $\beta_0\lesssim10^3$. The fluctuation level increases from the disk
midplane, and reaches order unity at disk surface as the disk become more and more
magnetically dominated. The case with $\beta_0=10^2$ show distinctive features from
higher $\beta_0$ cases: except for a small bump around the disk midplane, the density
fluctuation level stays roughly constant at about $0.5$. We note that the turbulent
kinetic energy profile in run B2 is similar to that in run B3, while the density near disk
surface in run B2 is a factor of several higher than run B3, indicating smaller turbulent
velocity, hence the profile of  $\langle\delta\rho^2\rangle^{1/2}/\rho$ for run B2 drops
below that for run B3, which is essentially a reflection of weaker turbulence for run B2.


\begin{figure}
    \centering
    \includegraphics[width=90mm]{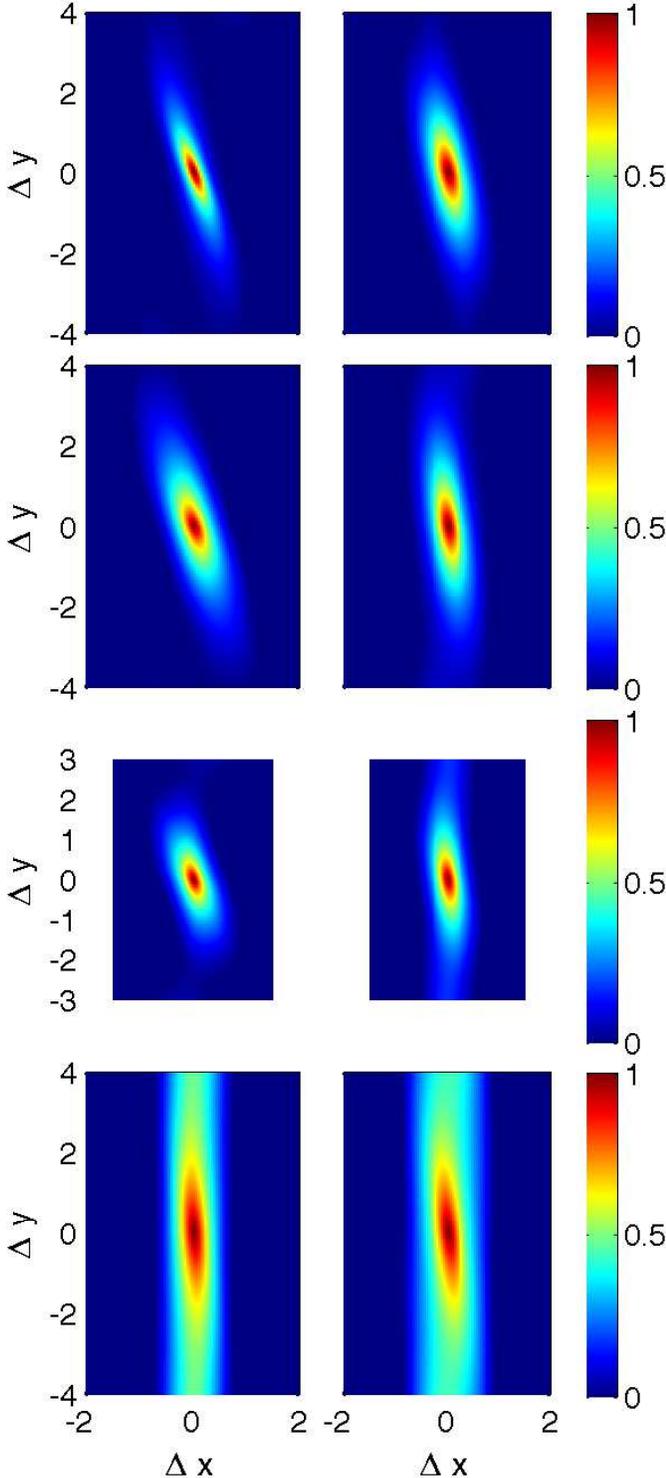}
  \caption{Two-dimensional autocorrelation function (ACF) in the horizontal plane for
  magnetic energy fluctuations for our simulations B4 ($\beta_0=10^4$), B3, B3-hr 
  ($\beta_0=10^3$) and B2 ($\beta_0=10^2$), from top to bottom.
  Left panels correspond to the ACFs for $|z|<2.5H$, while right panels are for
  $|z|>2.5H$. Note that the domain size for run B3-hr is slightly smaller.}\label{fig:acf}
\end{figure}

\subsubsection[]{Autocorrelation Function}

Another useful diagnostic of the MRI turbulence is the autocorrelation function (ACF)
of kinetic and magnetic energy fluctuations, which is defined as
\citep{Guan_etal09,Simon_etal12a}
\begin{equation}
{\rm ACF}(E_{K, B}, \Delta{\mb x})=\bigg\langle
\frac{\int\delta{\mb f}(t, {\mb x})\cdot\delta{\mb f}(t, {\mb x}+\Delta\mb{x})d^3{\mb x}}
{\int \delta f^2(t, {\mb x})d{\mb x}^3}\bigg\rangle_t\ ,
\end{equation}
where $f$ denotes ${\mb B}$ for magnetic energy, and $\sqrt{\rho}{\mb v}$ for kinetic
energy, the angle bracket denotes time average. Here we consider a two-dimensional
ACF in the horizontal plane at some fixed vertical height $z$, and we subtract the
horizontally averaged quantities in the evaluation of $\delta{\mb f}$:
$\delta{\mb f}\equiv{\mb f}(x,y,z)-\bar{\mb f}(z)$.
This is necessary as the mean flow and mean field can become particularly strong
in the presence of net vertical magnetic field.

In Figure \ref{fig:acf}, we show the ACFs for magnetic energy fluctuations in all our
simulations B2, B3, B3-hr and B4, for both near the midplane region (left, for
$z\leq2.5H$) and the disk surface (right, $2.5H\leq z\leq5H$). The ACFs for the
kinetic energy exhibits similar patterns, hence we do not involve a separate figure to
show them. The ACF for run B4 at midplane where the net vertical magnetic flux
is relatively small ($\beta_0=10^4$) is very similar to that found in zero net-flux
simulations \citep{Guan_etal09,Simon_etal12a}, containing a narrow and elongated
centroid that is tilted with respect to the azimuthal axis by about $\zeta_B\sim15^\circ$.
Such a tilt angle is related to the empirical correlation between the plasma $\beta$
and the stress parameter $\alpha\approx1/2\beta$ \citep{Guan_etal09,BaiStone11},
which also applies for the midplane region of our run B4 as we can compare Figures
\ref{fig:profile} and \ref{fig:amt}. Moving up to the disk surface, we find that the
centroid becomes broader in the upper layer, and the tilt angle also becomes smaller.
These features indicate longer correlation length in the radial direction, as well as
more isotropy in the magnetically dominated disk corona.
The ACFs in run B3 at the midplane has similar tilt angle as in run B4, typical for
MRI turbulence, while at both midplane and corona, the peaks in the ACFs are
as broad as the corona region of run B4, which is in line with the fact that the entire
disk has become magnetically dominated.

Our run B2 with $\beta_0=100$ exhibits an extreme situation: fluctuations are
highly elongated in the azimuthal direction, with the measured correlation length
comparable to the azimuthal size of our simulation box, which might suggest that
our azimuthal box size is not sufficient to properly resolve the MRI turbulence.
Nevertheless, the shape of the ACF indicates that the fluctuations are
quasi-axisymmetric (which is also evident as one views the raw simulation data),
and the radial fluctuations are well fitted into our simulation box.

\subsubsection[]{Numerical Convergence}

Due to the large computational cost, we have not carried out a full resolution
study in this paper. However, we do expect our simulation results to converge
in runs B2 and B3-hr where we have adequate resolution to properly resolve
the linear MRI modes as suggested by \citet{Latter_etal10}. Based on the
quality factor criterion \citep{Hawley_etal11}, we expect all our simulations to
well resolve the MRI turbulence. Below we compare the results between our
simulation runs B3 and B3-hr and briefly discuss about numerical convergence.

We see from Figure \ref{fig:acf} that the shape of the ACFs in our runs B3-hr
and B3 are similar, but the ACF in run B3-hr is more centrally-peaked than its
low-resolution counterpart. Since the ACF is simply the Fourier transform of
the power spectrum, this indicates that the high-resolution run gives a flatter
power spectrum with more power on the small scales. In addition, comparing
various profiles in Figure \ref{fig:profile} between runs B3 and B3-hr, we see
again that the profiles are very similar, with higher resolution giving slightly
higher turbulent energy. This is also reflected in Table \ref{tab:amt} where the
Shakura-Sunyaev $\alpha$ from run B3-hr is about $25\%$ higher. Similarly,
in Figure \ref{fig:drho}, we see that run B3-hr leads to slightly larger density
fluctuations. Later in Figure \ref{fig:wind}, slightly higher mass outflow rate is
achieved in run B3-hr. 

The systematically higher saturation amplitude of the MRI in the high-resolution
run discussed above might suggest that our simulations for $\beta_0=10^3$
have not fully reached numerical convergence. However, such higher
saturation amplitude may not be due to under-resolved MRI turbulence,
and several other factors are
likely to play a role. While slightly smaller horizontal box size in run B3-hr can
be a potential cause, the fact that the peaks of the ACF are well contained in
the box makes it somewhat unlikely \citep{Simon_etal12a}. Another possibility
is that in stratified MRI simulations, the property of the MRI turbulence is
largely affected by the dynamo process (see detailed study in Section
\ref{ssec:dynamo}), which generates strong mean toroidal field that changes
sign over time. It appears that the instantaneous large-scale toroidal field
generated in run B3-hr is on average stronger than that in run B3, as can be
judged from Figure \ref{fig:dynamo}, and both are above equipartition field
strength. Such larger mean toroidal field leads to stronger magnetic support to
the disk, resulting in slightly different vertical disk structures between the two
runs B3 and B3-hr. In this respect, one should be careful when discussing
the issue of convergence, since comparing turbulent properties between
simulations with different disk structures can be misleading. Since the
mechanism of the dynamo is far from being understood, and the dynamo in
the presence of strong net vertical flux behaves very differently from the case
with zero net vertical flux, we defer this convergence issue for future study.

In sum, although quantitative differences exist between runs B3 and B3-hr,
the physical properties of the MRI turbulent disk from the two runs all agree
with each other. We therefore consider both runs as valid representations of
the physical system, with run B3-hr as a more accurate model.

\subsection[]{Energetics and Angular Momentum Transport}\label{ssec:amt}

There are two sources of angular momentum transport in accretion disks,
namely, the radial transport (or re-distribution) via turbulence and the vertical
transport via disk wind. We study them separately in this subsection and
then briefly discuss about the energetics in our simulations.

\subsubsection[]{Radial Transport of Angular Momentum}

\begin{figure}
    \centering
    \includegraphics[width=85mm]{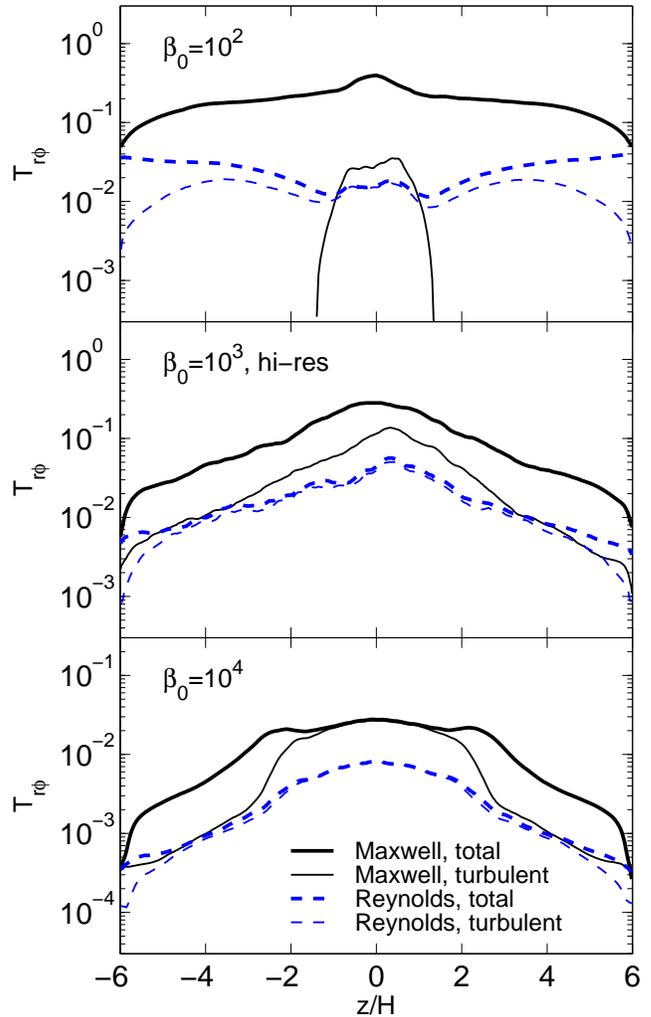}
  \caption{Vertical profiles of Maxwell stress (solid) and Reynolds stress (dashed)
  for runs B2, B3-hr and B4 (from top to bottom panels). In each panel, we show
  the total stress (bold) as well as contribution from turbulent fluctuations (thin).
  The stresses are normalized by $\rho_0c_s^2$.}\label{fig:amt}
\end{figure}

The radial transport of angular momentum is characterized by the $\alpha$
parameters defined in equation (\ref{eq:alpha2}). In Figure \ref{fig:amt}, we show
the vertical profiles of Maxwell and Reynolds stresses for all our simulations. Both
the Maxwell and Reynolds stresses increases monotonically with net vertical
magnetic flux, and approaches a few times $0.1\rho_0c_s^2$ for $\beta_0=10^2$.
In the mean time, the profile gradually changes from centrally peaked
for large $\beta_0$, to a more or less flat profile at $\beta_0=10^2$. In particular, the
slope of the wing beyond $\pm2H$ becomes more level as $\beta_0$ decreases. We
can further compare our run B4 with zero net vertical flux stratified MRI simulations
\citep{SHGB96,MillerStone00,GuanGammie11,Simon_etal12a}, where the slope of the
wing is further steeper. These results suggest a general trend that the surface region
plays a more and more important role in the radial angular momentum transport as the
net vertical magnetic field increases. Also, we see that the Maxwell stress is always
larger than the Reynolds stress, but the Reynolds stress becomes more and more
important toward disk surface.

To further decode the mechanism for the radial transport of angular momentum, we
separate the contributions from the mean field/mean flow
($\langle\bar{\rho}\bar{v}_x\bar{v}'_y\rangle$ or Reynolds stress and
$\langle-\bar{B}_x\bar{B}_y\rangle$ for Maxwell stress) and the turbulent
stress (the rest). We remind the reader that we compute the mean stress from
horizontally averaged quantities in each snapshot and then take the time average.
As net vertical field increases, the turbulent component increases
steadily with net field and saturates for $\beta\lesssim10^3$ in a way similar to the
turbulent energy density profile in Figure \ref{fig:profile}. There are two intriguing
features Figure \ref{fig:amt}. First, in the disk surface region (near our vertical
boundary), the stress is dominated by the mean field/mean flow components (i.e.,
magnetic breaking) rather than turbulence, and the contribution from the mean
field/mean flow components rapidly increases as net field increases. In run B4, the
mean field component dominates the wings of the Maxwell stress, while in runs
B3-hr (and B3), the mean field component dominates the entire disk. Second, for
sufficiently strong background magnetic field $\beta_0\lesssim10^2$, the turbulent
stress become completely unimportant at all locations, and the turbulent Maxwell
stress can even become negative. Interestingly, the central region where the
turbulent Maxwell stress is positive coincides with the bump in the density fluctuation
(Figure \ref{fig:drho}).

The decline of the turbulent component and rise of the mean field/mean flow
component of the stress as one increases net vertical flux strongly contrast with
conventional unstratified shearing-box simulations. In those simulations, the initial
net toroidal magnetic field is mostly set to zero. Although the net toroidal
magnetic flux is a not conserved quantity, it generally stays very close to zero for
the duration of the simulations. With stratification, the large-scale toroidal field
evolves substantially and for $\beta_0\lesssim10^3$, it well exceeds equipartition
strength (see Figures \ref{fig:dynamo} and \ref{fig:bt100prof}). Although the linear
growth of the MRI is not directly coupled to the toroidal field, the non-linear
saturation certainly depends on it \citep{HGB95}. Therefore, it is not appropriate
to compare unstratified MRI simulations with stratified simulations with the same
$\beta_0$ unless the unstratified simulation contains a substantial net toroidal
magnetic field.

\begin{table*}
\caption{Energetics and Angular Momentum Transport.}\label{tab:amt}
\begin{center}
\begin{tabular}{ccccccccc}\hline\hline
 Run & $\langle\alpha_{\rm Max}\rangle$ & $\langle\alpha_{\rm Rey}\rangle$
 & $\langle T_{z\phi}^{\rm Max}\rangle$ & $\langle T_{z\phi}^{\rm Rey}\rangle$
 & $P_{\rm sh}$ & $\dot{E}_K$ & $\dot{E}_P$ & $\dot{m}_w$\\\hline
B2 & 0.92 & 0.12 & 0.061 & 0.048 & 3.90 & 0.89 & 1.26 & 0.0944 \\
B3 & 0.37 & 0.077 & 0.0085 & 0.0041 & 1.69 & 0.079 & 0.15 & 0.0138 \\
B3-hr & 0.47 & 0.086 & 0.010 & 0.0046 & 2.08 & 0.086 & 0.21 & 0.0160 \\
B4 & 0.061 & 0.014 & 0.00072 & 0.00036 & 0.28 & 0.0065 & 0.010 & 0.0013\\
\hline\hline
\end{tabular}
\end{center}
Note: All in natural units, with $\rho_0=c_s=\Omega=1$.
\end{table*}

Integrating the stresses one obtain the $\alpha$ parameter defined in (\ref{eq:alpha2}),
and the results are listed in Table \ref{tab:amt}.
The $\alpha$ parameter increases monotonically with net
magnetic flux, from a total of about $\langle\alpha\rangle\approx0.075$ for
$\beta_0=10^4$ to $\langle\alpha\rangle\approx1.0$ for $\beta_0=100$,
The ratio of $\langle\alpha_{\rm Max}\rangle$ to
$\langle\alpha_{\rm Rey}\rangle$ increases from about 4 for $\beta_0=10^4$, which
is similar to that in zero net vertical flux simulations (e.g., \citealp{Davis_etal10}), to
about 7 for $\beta_0=10^2$, which is mainly due to an increasing contribution from
the mean (rather than turbulent) magnetic field.

We note that in run B2, the Maxwell stress only decreases very slowly with height, while
the Reynolds stress even increases with height, making the volume integrated stress
limited by the vertical extent of our simulation box. However, we argue that increasing the
box size does not necessarily improve the situation. In this case, the radial angular
momentum transport is dominated by the mean field and mean flow. As we shall discuss
in Section \ref{sec:wind}, the strength of the large-scale mean field at disk surface is
intimately connected to the global condition of the disk and can not be determined in the
local shearing-box approch. Therefore, uncertainty still remains as one uses a taller
box. The measured $\langle\alpha\rangle$ value for run B2 should be only taken as a
reference.

\subsubsection[]{Vertical Transport of Angular Momentum}

Angular momentum can also be extracted from the disk directly in the vertical direction
through outflow and magnetic fields, which is determined by the $z\phi$ components
of the Reynolds and Maxwell stresses exerted at vertical boundaries
\begin{equation}
\begin{split}\label{eq:windT}
T_{z\phi}^{\rm Rey}&=(\rho v_yv_z)|_{\rm bot, top}\ ,\\
T_{z\phi}^{\rm Max}&=(-B_yB_z)|_{\rm bot, top}\ ,\\
\end{split}
\end{equation}
where in practice subscripts $_{\rm bot}$ and $_{\rm top}$ mean that we measure these
quantities at the last layer of the top and bottom of the computational domain. The torque
exerted by the wind stresses can be obtained by simply multiplying
$\langle T_{z\phi}^{\rm Rey}\rangle$ and $\langle T_{z\phi}^{\rm Max}\rangle$ by the
radius $R$. Including the contributions from radial and vertical transports, the accretion
rate can be approximately written as\footnote{With the assumption that the vertical
integral of $T_{r\phi}$ is independent of radius. This corresponds to Equation (18) of
\citet{Fromang_etal12} with $p+q+2=1/2$.}
\begin{equation}
\dot{M}=\dot{M}_r+\dot{M}_z\approx\frac{2\pi}{\Omega}\int dz T_{r\phi}
+\frac{4\pi R}{\Omega}T_{z\phi}\bigg|^{\infty}_{-\infty}\ ,\label{eq:mdot}
\end{equation}
where $\dot{M}_r$ and $\dot{M}_z$ represent contributions to the accretion rate from
radial and vertical transport respectively.
Qualitatively, one may replace the vertical integral $\int dz$ by the disk scale height $H$.
We see that given the same level of stress $T_{r\phi}$ and $T_{z\phi}$, vertical transport
is more efficient than radial transport by a factor of about $R/H$.

Being a local shearing-box simulation, however, $R$ is unspecified, and moreover,
the location of the central object is unspecified (can either at inner or outer $\hat{\mb x}$
direction), making the sign of $R$ ambiguous. A physical disk wind requires that the sign
of $T_{z\phi}$ on the two vertical boundaries be the opposite and are steady, and is
closely related to the symmetry of the wind solution as we will discuss in detail in Section
\ref{ssec:fate}. In this subsection, we only consider the absolute value of $T_{z\phi}$
and average it over the two vertical boundaries (i.e., assuming the flow structure and
magnetic configuration has the physical geometry)\footnote{We will discuss in Section
\ref{ssec:fate} that the outflow is unlikely to be {\it directly} connected to an ordered
disk wind mainly due to geometric reasons.
Let us set aside the issue with outflow geometry here for the purpose of discussion.},
hence the relative importance between radial and vertical transport can be rewritten as
\begin{equation}
\frac{\dot{M}_r}{\dot{M}_z}=\frac{\sqrt{2\pi}}{4}\frac{H}{R}\frac{\alpha}
{|T_{z\phi}^{\pm\infty}|/\rho_0c_s^2}\ .\label{eq:mdot1}
\end{equation}

We list the values of $T_{z\phi}^{\rm Rey}$ and $T_{z\phi}^{\rm Max}$ from each
simulation run in Table \ref{tab:amt}, with both normalized by $\rho_0c_s^2$.
The wind stresses appear to depend more sensitively on $\beta_0$, and increase by
about two orders of magnitude from $\beta=10^4$ to $\beta=100$. This may suggest the
importance of wind transport relatively to turbulent transport increases with net vertical
field. Taking $H/R=0.1$, we find that wind transport, if present,
should be small  (less than $25\%$) compared with radial transport for $\beta_0=10^4$,
which is consistent with the results by \citet{Fromang_etal12}. It would become more or
less comparable to radial transport for $\beta_0=10^3$, while it would dominate radial
transport for $\beta_0=10^2$. In addition, the contributions from Maxwell
$T_{z\phi}^{\rm Max}$ and Reynolds $T_{z\phi}^{\rm Rey}$ components are roughly
$2:1$ in runs B3 and B4, while the Reynolds component becomes more important for
run B2. However again, as we shall discuss in Section \ref{sec:wind}, the wind stress
can only determined from global approach, while our measured $\langle T_{z\phi}\rangle$
values should also be taken as a reference.

\subsubsection[]{Energetics}\label{sssec:energetics}

Using an isothermal equation of state, total energy is not conserved in the simulation.
However, it is important to examine the energy budget for consistency such that the work
done by the Keplerian shear (i.e., from radial shearing-box boundaries) exceeds the energy
loss from the vertical boundaries, with the rest of the energy presumably escape from the
disk in the form of radiation.

In ideal MHD, the energy flux is given by
\begin{equation}
{\mb F}_E=\bigg(\frac{1}{2}\rho u^2+\epsilon+P_{\rm gas}+B^2\bigg){\mb u}
-({\mb B}\cdot{\mb u}){\mb B}\ ,
\end{equation}
where $\epsilon$ is the internal energy. Consider the net energy flux entering the two sides
of the radial boundaries, where we compute ${\mb F}_E\cdot{\mb e}_x$ and integrate
over height and azimuth. Note that ${\mb u}$ contains background shear, and one should
shift $u_y$ by $3\Omega L_x/2$ between the two boundaries. Correspondingly, only terms
containing ${\mb u}$ contribute, while all other terms cancel due to the shear-periodic
boundary condition. The non-vanishing terms turn out to be Maxwell and Reynolds stress,
and the work done by the radial boundaries per unit horizontal area per unit time read
\begin{equation}
P_{\rm sh}=\frac{1}{L_xL_y}\int\frac{3}{2}\langle \rho v_xv_y-B_xB_y\rangle\Omega
L_xdydz=\frac{3}{2}\Sigma c_s^2\Omega\langle\alpha\rangle\ ,
\end{equation}
where $\Sigma=\int\langle\bar{\rho}(z)\rangle dz$ is the column density.

The energy loss from vertical boundaries is given by
\begin{equation}\label{eq:Edot}
\dot{E}=\dot{E}_K+\dot{E}_P=\bigg[\rho\bigg(\frac{v^2}{2}+c_s^2\bigg){\mb v}
-({\mb v}\times{\mb B})\times{\mb B}\bigg]\cdot{\mb n}\bigg|_{\rm top+bot}\ ,
\end{equation}
where ${\mb n}$ is the unit vector pointing away from the disk in the vertical direction,
and we sum over contributions from the top and bottom vertical boundaries.
The first terms in the bracket represent the kinetic energy loss and the $PdV$ work done
by the mass outflow, and the second term represents energy loss from the Poynting flux
\footnote{In principle, one should use the total velocity ${\mb u}$ rather than ${\mb v}$
in Equation (\ref{eq:Edot}). However, linear terms that contain background shear should
vanish over space and time average. The quadratic term that contains background shear
corresponds to the kinetic energy of the Keplerian profile, which is compensated exactly
by the addition of mass to the simulation domain.}.
We have ignored the energy loss term associated with internal energy due to the mass
outflow, since we keep feeding mass to the system which balances this term exactly.
This term is comparable to the $PdV$ work term which, as we shall see, is always small
compared with the total $\dot{E}$ (hence our mass addition do not substantially
alter the energy balance).

In Table \ref{tab:amt} we also list the values of $P_{\rm sh}$, $\dot{E}_K$ and $\dot{E}_P$,
normalized in natural units. We see that $P_{\rm sh}$ is always larger than the sum of
$\dot{E}_K$ and $\dot{E}_P$, hence energy conservation is not violated. Also, we see
that $\dot{E}_P$ is always larger than $\dot{E}_K$ by a factor of about $1.5$ to $2$, hence
Poynting flux dominates energy loss. In addition, under the assumption of isothermal
equation of state adopted in our simulations, we find that as net vertical magnetic field
increases, $\dot{E}_K+\dot{E}_P$ roughly increases as $1/\beta_0$, which is much faster
than the rate $P_{\rm sh}$ increases. This may suggest radiative efficiency decreases
as a larger fraction of work done by the shear is converted into the wind energy loss.

\subsection[]{Dynamo Activities}\label{ssec:dynamo}

\begin{figure*}
    \centering
    \includegraphics[width=180mm]{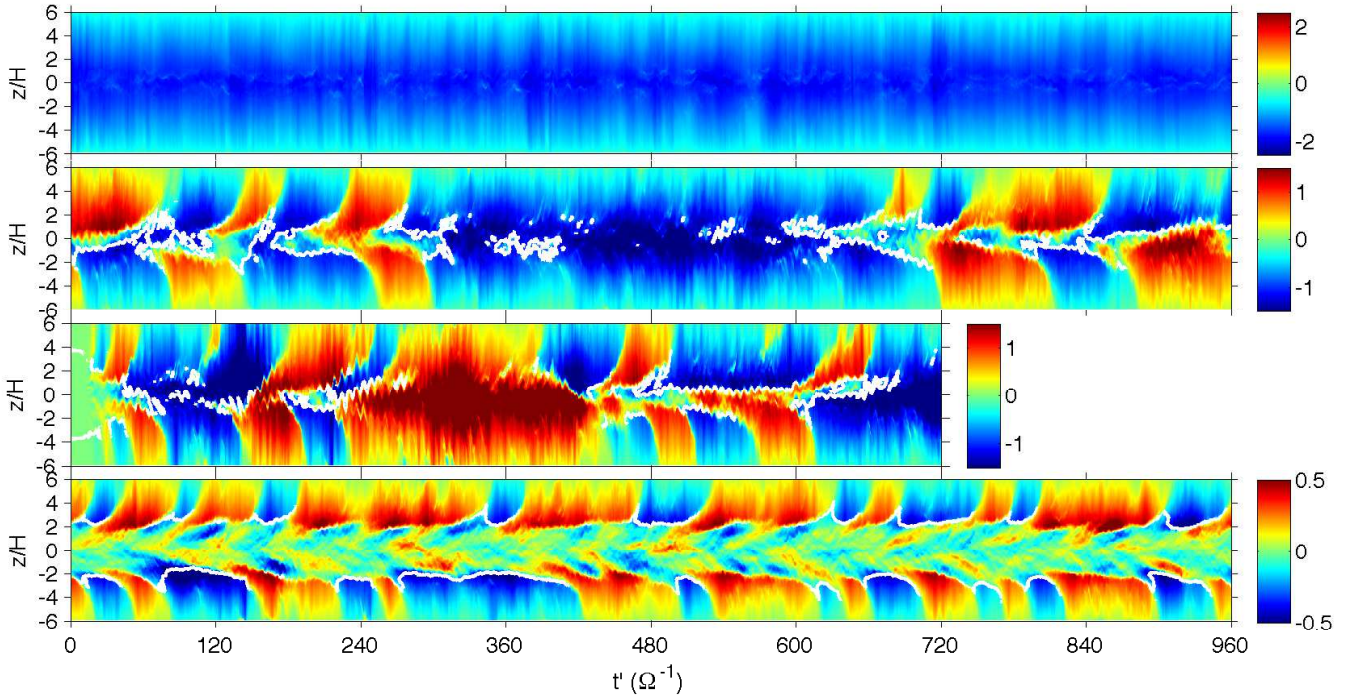}
  \caption{The time evolution of the horizontally averaged profiles of toroidal magnetic field
  in all simulation runs, from top to bottom: B2, B3, B3-hr and B4. Here we mark $t'=0$
  as the time when full net vertical flux is added ($t=240\Omega^{-1}$ for B2, B3 and B4,
  $t=0$ for B3-hr). The white contours mark plasma $\beta=1$.}\label{fig:dynamo}
\end{figure*}

It is well known in stratified shearing-box simulations with zero net vertical magnetic flux
that the mean toroidal field component experiences periodic flips on the time scale of $10$
orbits (e.g., \citealp{Brandenburg_etal95,Shi_etal10,Davis_etal10,Simon_etal12a}). Since
a zero background vertical field does not give any instability on its own in the linear theory,
the MRI activity in these simulations together with the periodic flips strongly suggest
that a magnetic dynamo is in operation. It is generally believed that shear, turbulence
and large scale azimuthal field are responsible for sustaining the dynamo cycles
\citep{VishniacBrandenburg97,Yousef_etal08,LesurOgilvie08,Gressel10}, yet the detailed
mechanism remains unsettled.

In the presence of net vertical magnetic field, the MRI can be sustained from the linear
modes hence a dynamo is not necessary to explain the turbulence. With very weak vertical
net magnetic flux, dynamo behavior was also observed in \citet{SuzukiInutsuka09},
\citet{Suzuki_etal10} and \citet{Fromang_etal12}, whose simulations correspond to
$\beta_0=10^4-10^6$, although it was not studied in detail.
Our simulations allow us to systematically study the evolution of the MRI dynamo with
net vertical magnetic flux. In Figure \ref{fig:dynamo}, we show the space-time plot of the
horizontally averaged azimuthal magnetic field for
all our simulation runs. It is obvious that for $\beta_0=10^4$, the mean azimuthal field
still undergoes cyclic oscillations between positive and negative signs, while
for $\beta_0=100$, the dynamo behavior disappears, where the mean azimuthal field
in the entire simulation box is completely dominated by one sign at all times. The case
with $\beta_0=10^3$ is somewhat marginal, where the dynamo-like behavior is present
for part of the time while for the rest of the time the mean azimuthal field is
predominantly one sign.

In zero net vertical flux stratified shearing-box simulations (see Figure 15 of
\citealp{Simon_etal12a} for a most clear rendering), the dynamo pattern is highly
repeatable in the space-time plot, also known as the ``butterfly diagram", with a
well-defined of period about 10 orbits. The dynamo cycles in our run B4 with
$\beta_0=10^4$, however, is highly irregular. Such behavior is also observed in
\citet{Fromang_etal12}. Further including the marginal case of run B3 with
$\beta_0=10^3$, we see that there is a systematic trend that as net vertical field
increases, the dynamo cycle weakens by becoming more irregular with less
periodicity. The flipping of mean azimuthal field becomes more sporadic, and the
mean time interval for field flipping also becomes longer. As $\beta_0$ reaches
below $10^3$, the flipping time interval virtually becomes infinity, and the
dynamo is completely quenched.

We note that the most unstable linear MRI mode is properly resolved in run B2,
which does not show dynamo behavior, but may not be well resolved in runs B3
and B4, which show dynamo activities. Hence question arises on whether the
dynamo activity depends on the proper resolution of the linear MRI modes. Our
run B3-hr, which is the same as run B3 but properly resolves the most unstable
MRI mode, is designed to clarify this potential ambiguity. We see that the
space-time pattern for the mean azimuthal field is very similar in the two runs:
dynamo activity appears only part of the time. This comparison further justifies
that the dynamo behaviors observed in our simulations are real.

It is natural to ask about what physical effects control the dynamo activities.
Phenomenologically, we see that in the presence of the dynamo, the disk is separated
into two distinct regions, namely the region near the midplane with relatively weak
magnetic field where the dynamo appears to be developed, and a highly magnetized
region outside. In Figure \ref{fig:dynamo}, we have overplotted white contours that
mark the total plasma $\beta=1$ (i.e., total magnetic pressure equals the gas pressure).
Remarkably, these white contours perfectly separate the two regions for both runs B3
and B4, while for run B2, the entire disk is magnetically dominated hence there is no
contour at all times. It is intriguing to notice that the dynamo is present whenever the
magnetic field strength in the disk midplane region is below equipartition strength.

Meanwhile, the criterion of $\beta=1$ also applies to magnetic buoyancy (i.e.,
interchange and Parker-type instabilities, \citealp{Newcomb61,Parker66}) as discussed
in a number of zero net-flux simulations
\citep{Blaes_etal07,Shi_etal10,GuanGammie11,Simon_etal11}.
With isothermal equation of state, and assuming zero mean vertical field, the fluid is
buoyantly unstable if the magnetic energy density decreases with height
\citep{GuanGammie11}. It turns out that the magnetic field strength tends to be
constant in the gas pressure dominated disk midplane regions, presumably due to
efficient turbulent mixing, while it falls off with height in the magnetically dominated
corona. Correspondingly, for regions with $\beta<1$, the fluid is unstable due to
magnetic buoyancy, while for regions with $\beta>1$, the fluid is buoyantly (marginally)
stable and dynamo activities produce cyclic alternations of the mean magnetic field.
Although the analysis of the Parker instability in previous works assumes zero mean
vertical field, it continues appear to be valid in our simulations with net vertical field,
which is likely because the net vertical field in our simulations is much weaker (by at
least a factor of $10$) than the mean azimuthal field.

The strong azimuthal magnetic field in our simulations with $\beta_0\lesssim10^3$ is to
some extent similar to the simulations performed by \citet{JohansenLevin08}. They initiated
their simulations by a pure azimuthal field with equipartition strength (with zero net vertical
magnetic flux), and found that the interplay between the Parker instability at disk surface
layer and the MRI near the midplane effectively produces a magnetic dynamo, and the
azimuthal flux is well confined to the disk. Simulations of \citet{Gaburov_etal12} for the tidal
disruption of a molecular cloud approximately realize the situation.
The key difference in our case is that a strong outflow is produced in the presence of net
vertical magnetic field (see Section \ref{sec:wind}),
hence azimuthal flux continuously escape the simulation box and has to be continuously
generated within the disk. Therefore, although Parker instability is likely to be present in the
entire disk when $\beta_0\lesssim10^3$, the dynamo mechanism of Johansen \& Levin no
longer operates in the presence of strong net vertical magnetic field. On the other hand,
magnetic dissipation due to the Parker instability is likely to play an important role on the
thermal structure of the disk \citep{Uzdensky12}, which deserves future exploration.

\section[]{Simulation Results: Outflow}\label{sec:wind}

It has been found in \citet{SuzukiInutsuka09} and \citet{Suzuki_etal10} that the
inclusion of a net vertical magnetic field leads to strong mass outflow in shearing-box
MRI simulations, and the rate of the mass outflow is approximately proportional to
$1/\beta_0$. The mass outflow was interpreted as the launching of a wind from the
disk and would potentially connect to a \citet{BlandfordPayne82} type global disk wind. 
The connection between MRI-driven outflow and global disk wind has been further
investigated by \citet{Fromang_etal12}, who focused on the case with $\beta_0=10^4$.
They argue that the MRI is able to launch the magnetocentrifugal wind that is strongly
time-dependent. Similar conclusion has been made by \citet{Lesur_etal12} for the case
with $\beta_0\sim10$, where there is only one single MRI mode fitted into the disk
with a physical symmetry for the wind, and secondary instabilities are likely to make it
time-dependent. Very strong outflows are also observed in our simulations. However,
we argue that the outflow seen in shearing-box MRI simulations is unlikely to be
directly connected to a global disk wind, as we elaborate below.

\subsection[]{Structure of the Outflow}\label{ssec:outflow}

\subsubsection[]{Critical Points}

The global disk wind model by \citet{BlandfordPayne82} states that when the magnetic
field at the surface of a thin disk is inclined relative to the disk normal for more than
$30^\circ$, outflowing gas can be accelerated centrifugally along field lines, extracting
angular momentum from the disk, and is gradually collimated by the magnetic hoop
stress. A crucial ingredient of this picture is the wind launching/mass loading from the
disk, which is intrinsically connected to the gas dynamics in the disks. The wind
launching process has been studied analytically under the local shearing-box
approximation \citep{WardleKoenigl93,OgilvieLivio01,Ogilvie12}, where all the radial
gradients except the shear are omitted to reduce the problem to fully one-dimensional
(1D). The 1D solutions are then matched to global wind solutions, where the mass
loading rate is determined as an eigen-value problem.

\begin{figure}
    \centering
    \includegraphics[width=75mm]{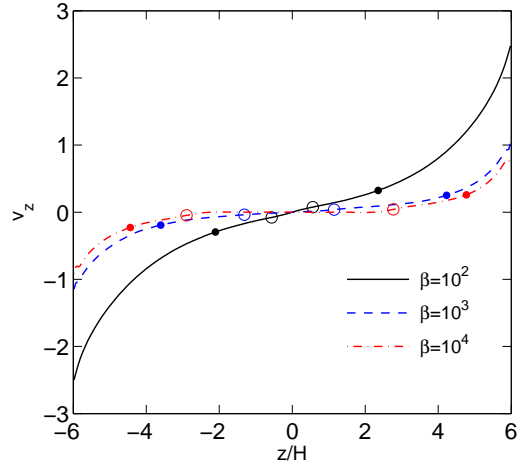}
  \caption{Horizontally-averaged vertical velocity profiles for fiducial resolution
  simulations B2, B3 and B4 with $\beta_0=10^2$ (solid), $10^3$ (dashed)
  $10^4$ (dash-dotted) respectively. Also marked are the locations of Alfv\'en points
  (solid dots) and slow magnetosonic points (circles). The fast magnetosonic points
  are beyond the simulation domain.}\label{fig:criticalpts}
\end{figure}

A physical wind solution should pass three critical points, namely, the slow and fast
magnetosonic points, and the Alfv\'en point. In the 1D (laminar) model, they are given
by
\begin{equation}\label{eq:fastslow}
v_z^2=\frac{1}{2}(c_s^2+v_A^2)\pm\frac{1}{2}\sqrt{(c_s^2+v_A^2)-4c_s^2v_{Az}^2}\ ,
\end{equation}
for fast and slow magnetosonic points, and
\begin{equation}
v_z=\pm v_{Az}\ ,\label{eq:Apoint}
\end{equation}
for the Alfv\'en point, where $v_{Az}=\sqrt{B_z^2/\rho}$ is the vertical component of the
Alfv\'en velocity, and $v_A$ is the full Alfv\'en velocity. The requirement that the flow
passes smoothly through these critical points poses three eigen-value problems that
would determine three physical quantities, such as the mass loading rate and the field
inclination at disk surface. In the local shearing-box approach, the fast magnetosonic
point, and sometimes the Alfv\'en point, are sufficiently high above the disk where the local
approximation no longer applies, and they need to be captured in the global models. The
lack of one or two critical points implies that the solution obtained within the computational
domain has one or two degrees of freedom.

In Figure \ref{fig:criticalpts}, we show the time and horizontally averaged
profiles of the gas vertical velocity as a function of height. In all cases, the vertical
velocity rapidly increases towards disk surface, and becomes transsonic or supersonic
as the flow leaves the simulation box. We also calculate the location of the critical points
based on equations (\ref{eq:fastslow}) and (\ref{eq:Apoint}), where the Alfv\'en velocity
is based on the time average of the absolute value of the horizontally averaged the
magnetic field.
We find that for all cases, the fast magnetosonic points are beyond our simulation box
(the fast magnetosonic speed is about $3c_s$ or larger at vertical boundaries),
while the Alfv\'en (marked in the Figure) and slow magnetosonic points are well contained
in the box. This fact means that the flow structure is not fully determined and has one
degree of freedom.

\subsubsection[]{Conservation Laws}\label{sssec:conslaw}

\begin{figure}
    \centering
    \includegraphics[width=75mm]{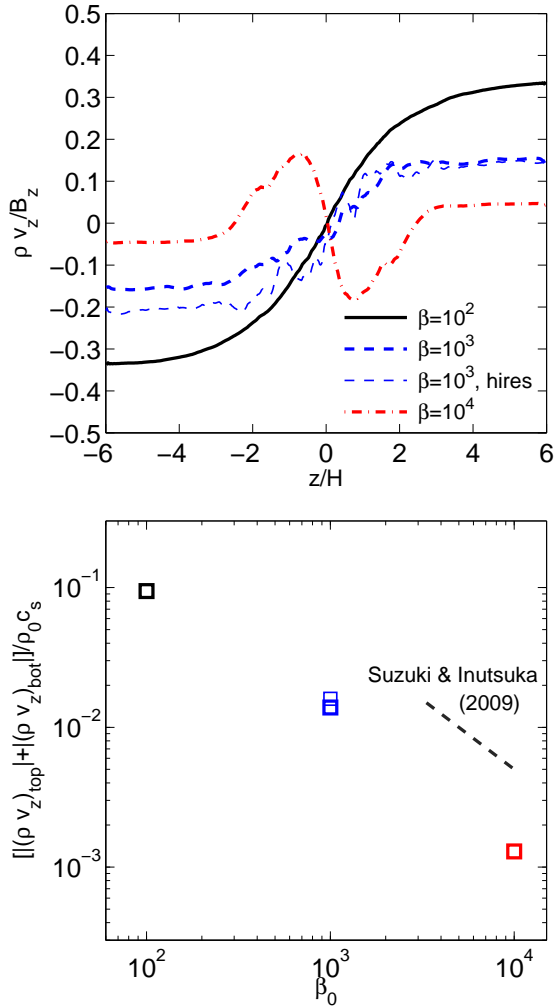}
  \caption{Top: time averaged vertical profile of mass outflow rate normalized to the
  mean vertical magnetic field (or $\kappa$, see Equation \ref{eq:chi}) for all our simulation
  runs. Bottom: time averaged mass loss rate from the simulation domain as a function
  of net vertical flux. Thick/thin symbols correspond to
  fiducial/high resolution simulations (B3-hr/B3).}\label{fig:wind}
\end{figure}

It is well known that for a laminar magnetized wind, the gas follows the magnetic field
lines, and the flow is characterized by a number of conserved quantities along the
streamlines \citep{BlandfordPayne82,PelletierPudritz92}. These conservation laws
provide very useful diagnostics that help us understand the mechanism for the
launching and acceleration of the outflow, as well as the angular momentum transfer
processes. The forms of these conservation laws in the shearing-box framework
are derived in \citet{Lesur_etal12}. In shearing-box, it was shown that poloidal gas
streamlines do not necessarily follow exactly the poloidal field lines, which brings new
terms to the conservation law. Nevertheless, the difference is generally small.
Since strong turbulence are present in all our simulations while these conservation
laws are derived for a laminar flow, they only hold approximately and serve for
diagnostic purpose.

The starting point is mass conservation, where
\begin{equation}
\overline{\rho v_z}={\rm const}\equiv \kappa\overline{B}_z\ ,\label{eq:chi}
\end{equation}
where overline indicates horizontal average, and $\kappa$ is constant. We show the
profile of $\overline{\rho v_z}/\overline{B}_z$ (or $\kappa$) for all our simulation runs in
the top panel of Figure \ref{fig:wind}. Because we constantly add mass to the
simulation box to maintain steady state, the vertical profile of $\overline{\rho v_z}$
does not become flat until beyond $z\sim\pm 3H$. There is some asymmetry in
the high-resolution run B3-hr, which is most likely due to its shorter run time and
the lack of statistics. The asymptotic values of $\overline{\rho v_z}$ are used to
calculate the mass loss rate $\dot{m}_w$ from the simulations
\begin{equation}
\dot{m}_w\equiv\langle\overline{\rho v_z}\rangle|_{\rm top}
-\langle\overline{\rho v_z}\rangle|_{\rm bot}\ ,
\end{equation}
and we have listed the value of $\dot{m}_w$ in Table \ref{tab:amt}. Further
details about the mass outflow rate will be discussed in Section \ref{ssec:mdot}.

As long as $\kappa$ is constant, as valid beyond $z\pm3H$ in our simulations,
specific angular momentum and energy are conserved along streamlines.
Following \citet{Lesur_etal12}, the specific angular momentum reads
\begin{equation}
f\equiv{\mc L}-\frac{\bar{B}_y}{\kappa}\ ,\label{eq:angcons}
\end{equation}
where ${\mc L}\equiv v_y+\Omega x/2$ is the fluid part of the specific angular
momentum. The partition between ${\mc L}$ and $\bar{B}_y/\kappa$ then describes
the angular momentum exchange between gas and magnetic field.

Energy conservation is given by the Bernoulli invariant, which reads
(\citealp{Lesur_etal12}, without correction terms)
\begin{equation}
\begin{split}\label{eq:bernoulli}
E_{\rm Ber}&=E_K+E_T+E_\phi+E_B\\
&=\frac{\bar{u}^2}{2}+c_0^2\log(\bar{\rho})+\phi-\frac{\bar{B}_yv_y^*}{\kappa}\ ,
\end{split}
\end{equation}
where $E_{\rm Ber}$ represents the specific energy along a streamline, with
the four terms denoting kinetic energy, enthalpy, potential energy and the work
done by the magnetic torque, respectively, and
\begin{equation}
v_y^*\equiv\bar{u}_y-\kappa\bar{B}_y/\bar{\rho}\ .
\end{equation}
Note that the full velocity ${\mb u}$ (rather than ${\mb v}$) enters $E_K$, and
the potential energy for the shearing-box is
\begin{equation}
\phi=-\frac{3}{2}\Omega x^2+\frac{1}{2}\Omega z^2\ .
\end{equation}

We extract the mean flow of the gas and integrate to obtain the streamline.
Because the disk midplane regions are generally the most turbulent with
$\kappa$ varying with height, we do not expect conservation laws to hold in
these regions region and only trace streamlines from $z=3H$ to $z=6H$,
setting $x=0$ at $z=3H$ (note that by construction the conservation laws
do not depend on the choice of the zero point of $x$).
This treatment covers the most interesting surface regions of the disk where
the launching and acceleration of the outflow take place. For illustration
purpose, we only consider run B2 and B4. For run B2, the large-scale flow
and magnetic structures are more or less steady (as seen from Figure
\ref{fig:dynamo}), hence we take the long-term average to obtain the
desired physical quantities along streamlines. For run B4, due to the
dynamo activities, we only pick up a short period between
$t'=560\Omega^{-1}$ and $t'=600\Omega^{-1}$, where the magnetic
structure in the upper side of the disk is quasi-steady, as judged from Figures
\ref{fig:dynamo} and \ref{fig:dynamo_bx}.

\begin{figure}
    \centering
    \includegraphics[width=85mm]{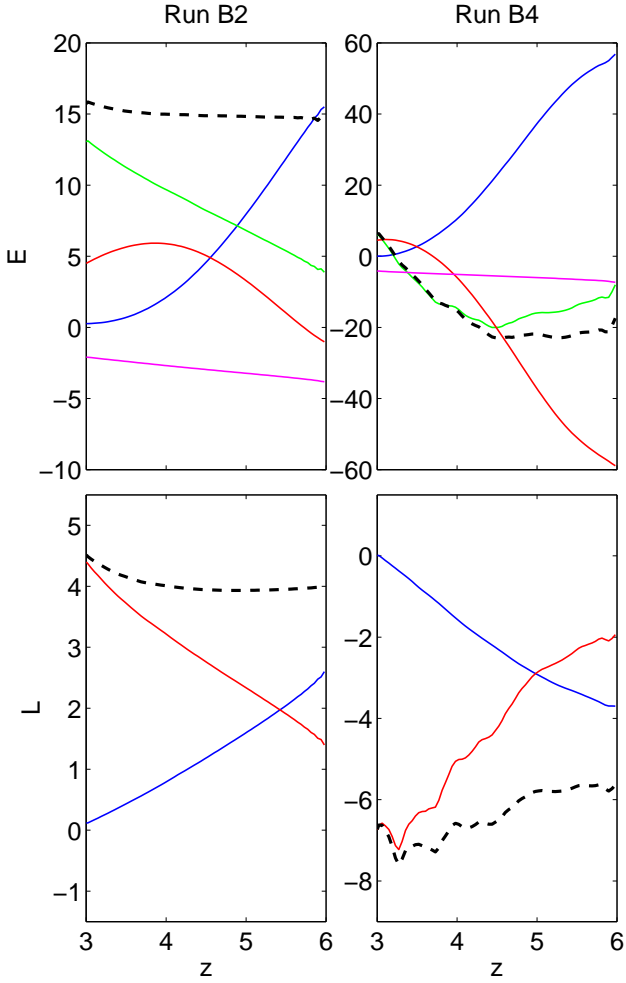}
  \caption{Energy (top) and angular momentum (bottom) along a streamline
  in the disk surface from $z=3H$ to $6H$ for run B2 (left, with long time average)
  and B4 (right, averaged between $t'=560\Omega^{-1}$ and $t'=600\Omega^{-1}$).
  For the energy plot, the various terms in Equation (\ref{eq:bernoulli}) are represented
  by: $E_{\rm Ber}$ (black dashed), $E_K$ (blue solid), $E_T$ (magenta solid),
  $E_B$ (green solid) and $\phi$ (red solid). For the angular momentum plot, the
  various terms in Equation (\ref{eq:angcons}) are represented by: $f$ (black dashed),
  ${\mc L}$ (blue solid), $-\bar{B}_y/\kappa$ (red solid).}\label{fig:conservation}
\end{figure}

In Figure \ref{fig:conservation} we show the vertical profiles of individual terms in
the specific energy and angular momentum for streamlines obtained from runs B2
and B4. We see from the black dashed lines that energy and angular momentum
are approximately conserved beyond about $4H$ for both cases. The rise of the
blue lines in the energy plots indicate flow acceleration. This is mostly compensated
by the reduction of potential energy (red) and work done by the magnetic torque
(green). Similarly, in the bottom panels of Figure \ref{fig:conservation}, we see that
the increase of fluid angular momentum is compensated by the reduction of magnetic
torque. Therefore, it becomes clear that the acceleration is magnetocentrifugal in
nature. Note that the role played by centrifugal potential $\phi$ and the magnetic
torque $-\bar{B}_yv_y^*/\kappa$ can be exchanged by shifting the zero point of $x$,
hence they represent the same effect. For example, if we were to trace the
streamlines from the disk midplane for the case of run B2, then we would found
that the acceleration is almost completely due to the centrifugal potential, since
for Run B2 the poloidal field lines are almost always inclined for more than
$45^\circ$ relative to disk normal, as can be seen in Figure \ref{fig:bt100prof}.

\subsection[]{Mass Loss Rate}\label{ssec:mdot}

In the bottom panel of Figure \ref{fig:wind}, we show the total mass outflow rate
$\dot{M}_w$ measured from the two vertical boundaries for all our simulations,
with the numbers given in Table \ref{tab:amt}. We see that $\dot{m}_w$ scales
roughly linearly with $1/\beta_0$, which is consistent with \citet{SuzukiInutsuka09}
and \citet{Suzuki_etal10}, while extending the range of $\beta_0$ from $10^4$
in their case down to $10^2$. From the definition of the Alfv\'en point, the mass
loss rate can be expressed as
\begin{equation}
\dot{m}_w=2\sqrt{\rho}\bigg|_A\cdot|B_z|\ ,
\end{equation}
where $|_A$ denotes value at the Alfv\'en point. A linear scaling of $\dot{m}_w$
with $1/\beta_0$ indicates that $\rho\propto B_z^2\propto 1/\beta_0$ at the
Alfv\'en point, which is in line with the fact that the location of the Alfv\'en points
decreases in height as net vertical field increases (the change in density profile
due to magnetic pressure support is not as significant). Moreover, based on the
same scaling, $v_z$ at the Alfv\'en point is expected to be constant, which is
roughly the case as seen from Figure \ref{fig:criticalpts}.

\subsubsection[]{Determining $\dot{m}_w$}

In Figure \ref{fig:wind}, we also indicate the scaling relation of $\dot{m}_w$
with $\beta_0$ from \citet{SuzukiInutsuka09}. We find that while our measured
$\dot{m}_w$ follows the same trend, while our proportional coefficient is about a
factor of $5$ smaller than theirs. Besides that our vertical box size is slightly
larger ($12H$ versus $8\sqrt{2}\approx11.3H$), this difference is mainly due to
the different outflow boundary conditions implemented in our simulations from
theirs. In fact, the rate of mass outflow $\dot{m}_w$ is not a well determined
quantity in local shearing-box type simulations because of the additional degree
of freedom. Moreover, in both \citet{Fromang_etal12} and \citet{Lesur_etal12},
it was found that $\dot{m}_w$ decreases as the vertical size of the simulation
box increases, which again suggests that $\dot{m}_w$ can not be reliably
obtained from shearing-box simulations.

\begin{figure*}
    \centering
    \includegraphics[width=180mm]{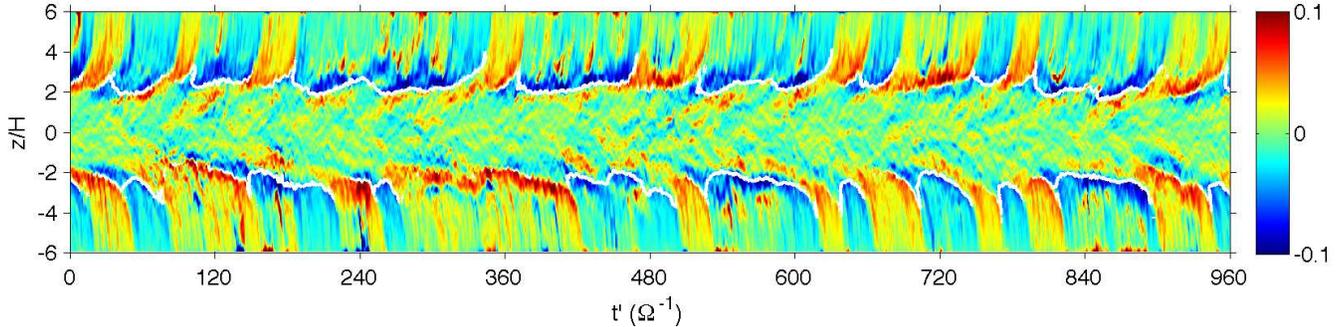}
  \caption{Same as Figure \ref{fig:dynamo}, but for run B4 only, and color represents
  the horizontally averaged radial magnetic field.}\label{fig:dynamo_bx}
\end{figure*}

Since the vertical gravity increases linearly with $z$ in shearing-box
approximation, the gas has to overcome stronger and stronger potential
well in order to flow out as box height increases. In reality, the vertical
increases only as $z/(R^2+z^2)^{3/2}$, where $R$ is the distance to the
central object, hence shearing-box approximation fails at $z\gtrsim R$.
\citet{Fromang_etal12} considered higher order expansion of the real
potential which would also require the radial potential to be modified to the
same order. They noticed that this procedure would introduce curvature
and no longer fit into the shearing-box framework.

The mass loss rate in real systems should be a well defined quantity,
and we speculate that it should be comparable to the mass loss rate in
shearing-box simulations performed with box size $L_z\approx2R$.
Suppose the mass outflow rate scales as
\begin{equation}
\frac{\dot{m}_w}{2\rho_0c_s}=\frac{A}{(L_z/H)^\nu}\frac{B_0^2}{2P_0}\ ,
\end{equation}
where $A$ is a dimensionless constant, the index $\nu$ describes the
scaling of $\dot{m}_w$ on the box height, and we have assumed that
$\dot{m}_w\propto1/\beta_0$ as suggested by the simulations discussed
in the previous subsection, with $B_0$ denoting the background vertical
field strength. With a little algebra, we find
\begin{equation}
\dot{m}_w=\frac{AB_0^2}{L_z\Omega}\bigg(\frac{L_z}{H}\bigg)^{1-\nu}\ .
\end{equation}
In particular, if $\nu\approx1$ (S. Fromang, 2011, private communication),
then our hypothesis ($L_z=2R$) yields
$\dot{m}_w\approx AB_0^2/2R\Omega$. This means that the mass loss
rate can be found once the coefficient $A$ is known. Moreover,
$\dot{m}_w$ is solely determined by the background
vertical field strength and does not depend on the disk thickness or the
disk surface density. On the other hand, if $\nu$ deviates from $1$, then
one would expect $\dot{m}_w\propto(H/R)^{\nu-1}$.

In sum, although $\dot{m}_w$ is not well determined from shearing-box
simulations, we argue that by carefully studying the dependence of
$\dot{m}_w$ on the height of the simulation box, it is possible to obtain
a physical estimate of the mass loss rate.



\subsubsection[]{Evolutionary Scenarios}

While we have demonstrated the correlation $\dot{m}_w\propto1/\beta_0$,
we have not discussed the range that it is applicable. Reading from Table
\ref{fig:amt}, we see that $\dot{m}_w$ for $\beta_0=100$ is somewhat smaller
than the expected linear correlation, which is suggestive of saturation. Indeed,
the rate of mass outflow measured in \citet{Lesur_etal12} for a similar
numerical setup (their run 1Dz6) with $\beta_0\sim10$ gives
$\dot{m}_w=0.234$ (multiplied by 2 to account for two surfaces), which is only
a factor of 2.5 times higher than our case with $\beta_0=100$
($\dot{m}_w=0.0944$). With further stronger magnetic field that is too strong
for the MRI to operate, \citet{Ogilvie12} showed that the rate of mass outflow
should rapidly decrease once $\beta_0$ drops below $1$. Therefore, the
scaling relation $\dot{m}_w\propto1/\beta_0$ holds up to $\beta_0\gtrsim100$,
beyond which the mass loss rate slows down and will eventually fall off
rapidly for strong super-equipartition field field.

These results, all together, suggest an evolutionary scenario. Assuming
no magnetic flux evolution, the disk would lose mass at constant rate
(assuming $\nu=1$) until the midplane net vertical field exceeds equipartition.
\citet{Lesur_etal12} explored such a scenario in their 1D simulations and
confirmed the steady decrease of mass loss rate with increasing
magnetization in the strong field regime. A strongly magnetized thin accretion
disk may gradually evolve into such a stage, where the rate of mass loss is
self-regulated so that it loses mass faster as more mass is fed into the disk
(increasing $\beta_0$, but still $\beta_0<1$), while loses mass slower if the
mass loss is not compensated (decreasing $\beta_0$). The disk structure
would be similar to a jet-emitting disk \citep{CombetFerreira08}, which is 
tenuous with super equipartition field.
Alternatively, if the evolution starts from a strongly magnetized thick disk,
the mass loss rate can be so large that the disk will be depleted in a few
orbits. For example, without feeding mass to our run B2, the mass loss time
scale is only about 5 orbits. Such rapid mass loss may lead to catastrophic
disruption of the disk, such as in core-collapse supernovae where rapidly
rotating progenitor core is threaded by extremely strong magnetic fields
\citep{Burrows_etal07}.




\subsection[]{Fate of the Outflow}\label{ssec:fate}

In this subsection we discuss whether the outflow launched from shearing-box
MRI simulations can be connected to a global Blandford-Payne type disk wind.
The key to this question lies in the symmetry. We note that due to the neglect of
curvature, shearing-box approximation does not specify which side of the radial
domain the central object is located. For a physical disk wind, symmetry requires
that the outflow from the top and bottom sides of the disk should be inclined
towards the same radial and azimuthal directions (which would be considered as
radially outward and trailing), and remains so at all times. Since gas flows along
magnetic field lines, the mean magnetic field at the top and bottom sides of the
disk should be bent to the same direction.

The first difficulty for outflow-wind connection is the magnetic dynamo, applicable
to the cases with $\beta_0\gtrsim10^3$ as we discussed in Secion \ref{ssec:dynamo}.
The constant flipping of the mean azimuthal (as well as radial, see Figure
\ref{fig:dynamo_bx}) field lines means that the direction of the outflow would
constantly swap between radially inward and outward, either of the directions would
be invalid for a global disk wind. Although our simulations with $\beta_0=10^4$ are
in many aspects very similar to simulations by \citet{SuzukiInutsuka09},
\citet{Suzuki_etal10}, and \citet{Fromang_etal12}, they argue that such phenomenon
leads to the time variability of the wind. However, we
stress here that the simple fact of cyclic sign change of the large-scale field is
already inconsistent with global wind geometry.

\begin{figure}
    \centering
    \includegraphics[width=90mm]{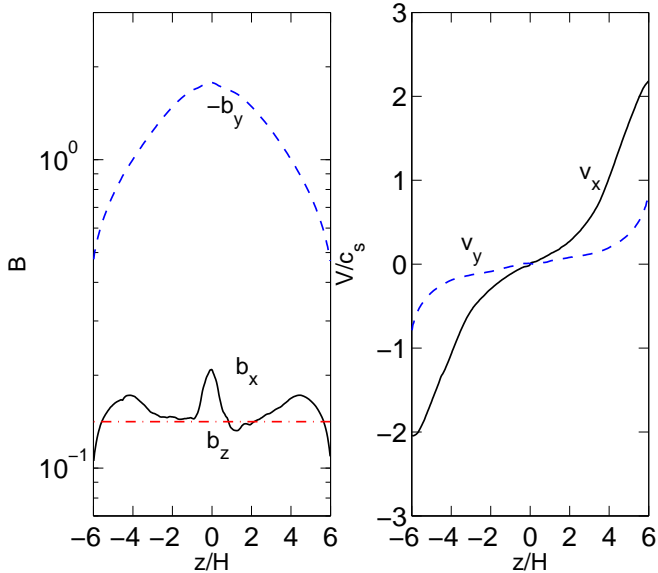}
  \caption{Time-averaged vertical profiles of magnetic field (left) and velocity (right)
  for run B2 with $\beta_0=100$. All in code units.}\label{fig:bt100prof}
\end{figure}

The second difficulty is that even there were no dynamo, such as the case for
$\beta_0=10^2$, the natural symmetry from the simulation is
unphysical for it to be connected to a global disk wind. In Figure \ref{fig:bt100prof},
we show the vertical profiles of the magnetic field and velocity fields for run B2.
Clearly, the outflows at the top and bottom sides of the disk are inclined toward
opposite directions, which is inconsistent with a global disk wind. Even in the
presence of the dynamo, there is a good chance that field lines at the top and
bottom sides tend to incline toward different directions (as can be tracked in
Figures \ref{fig:dynamo} and \ref{fig:dynamo_bx}).


Symmetry is also directly related to the vertical angular momentum transport by the
outflow, as we discussed in Section \ref{ssec:amt}. With the wrong symmetry, the wind
stress (\ref{eq:windT}) at the top and bottom sides of the disk have the same sign,
and cancel each other, hence the net vertical angular momentum transport vanishes.

One might attribute this ``wrong-symmetry" problem to the one extra degree of freedom in
shearing-box simulations, which could possibly be avoided by performing global simulations.
While it is certainly important to explore the same problem in global geometry, it has been
suggested by \citet{Ogilvie12} that global effects can be effectively taken into account by
applying some additional constraints on the vertical boundaries. Ogilvie constructed a 1D
model for launching disk wind / jet in the shearing-box approximation, where a much
stronger net vertical field $\beta_0\leq1$ was assumed so as to suppress the MRI. He
assigned the two free parameters at the vertical boundaries to be the strength of the radial
and azimuthal magnetic fields (he has two parameters because the Alfv\'en point turns out
to be beyond his computational domain), and was able to successfully construct laminar
wind solutions with the desired symmetry for the given constraints.

In view of this possibility, we further modify our vertical boundary conditions so that the
mean radial (or azimuthal) magnetic field in the last four grid zones at the vertical boundary
are set to some fixed value, which is done by first calculating the mean radial (or azimuthal)
field in these zones, and then subtracting a constant radial (or azimuthal) field uniformly in
all these cells to make the mean field equal to the desired value without violating the
divergence-free condition. As we see in Figure \ref{fig:bt100prof}, the mean radial field at
vertical boundaries is roughly the same as the mean vertical field in strength. Therefore,
we demand that the field incline by $45^\circ$ at vertical boundaries in the $\hat{x}-\hat{z}$
plane, and require that the mean radial fields at the two vertical boundaries have opposite
signs which conforms to the desired symmetry. However, after performing various
experiments on smaller domain test runs with parameters similar to runs B2, B3 and B4,
we find that such a treatment does not have an evident physical effect on the profile of
the mean flow and mean field, except for introducing a strong current sheet at the vertical
boundaries. This fact has also been discussed in \citet{Lesur_etal12}, and we conclude
that the poloidal field inclination at disk surface, particularly the symmetry of the mean
horizontal fields, is insensitive to the imposed vertical boundary condition the once the
Alfv\'en point is contained within the simulation domain.

In sum, we conclude that whether or not the MRI turbulence is accompanied by dynamo
activities, the outflow from the shearing-box MRI simulations is unlikely to be {\it directly}
connected to a global disk wind. The fate of the outflow remains to be explored using
global simulations.

\section[]{Discussion and Conclusions}\label{sec:conclusion}

\subsection[]{Summary}

In this paper, we have successfully performed local stratified shearing-box simulations
of the MRI that include a strong vertical magnetic flux with midplane plasma $\beta$
of the net vertical field ranging from $\beta_0=10^4$ to $10^2$, a regime that has not
been explored while is very likely relevant to accretion disks in many astrophysical
systems. Such a magnetic configuration gives rise to very vigorous MRI turbulence,
and simultaneously launches an outflow. We studied the properties of the MRI
turbulence as well as the disk outflow in detail and our major findings are summarized
below.

For the properties of the MRI turbulence, we find
\begin{itemize}
\item With relatively weak net vertical field of $\beta_0\gtrsim10^3$, the disk
consists of a gas pressure dominated disk midplane and a magnetic dominated
disk corona ($\beta\sim0.1-1$). By contrast, the entire disk becomes magnetically
dominated when $\beta_0\lesssim10^3$. The strong magnetic support substantially
modifies the vertical structure of the disk and makes it substantially thicker.

\item Turbulent magnetic and kinetic energies increase monotonically with
net vertical magnetic flux, and saturate when $\beta_0\lesssim10^3$. The MRI
also generates large scale toroidal magnetic field whose strength increases
monotonically with net vertical field, and dominates the total magnetic energy
over contribution from turbulence for $\beta\lesssim10^3$.

\item The Shakura-Sunyaev parameter $\alpha$ increases monotonically with
net vertical magnetic flux, ranging from $\alpha\sim0.08$ at $\beta_0=10^4$ to
$\alpha\gtrsim1.0$ for $\beta=10^2$. Maxwell stress dominates Reynolds stress
by a factor of 4-7. For weak net vertical flux of $\beta_0\gtrsim10^3$, turbulent
fluctuations dominates the contribution to $\alpha$, while for
$\beta_0\lesssim10^3$, radial transport of angular momentum is
dominated by the large scale fields.

\item The MRI dynamo that generates cyclic flips of the mean toroidal field
persists in the presence of weak net vertical magnetic flux, but becomes
more sporadic with less periodicity as net flux increases. The dynamo is
completely suppressed when the net flux is strong with $\beta_0<10^3$,
where the mean toroidal field exceeds equipartition strength and never flips.
\end{itemize}

Our results demonstrate the crucial dependence of the behavior of the MRI
turbulence on the net vertical magnetic flux. In particular, there is a critical net
vertical magnetic flux of $\beta_0\approx10^3$ near which many aspects of the
MRI turbulence change qualitatively. Additionally, more careful convergence
study is still needed, and
deeper understanding about the properties with MRI turbulence would further
benefit from the study of Prantl number dependence
\citep{Fromang_etal07,LongarettiLesur10,Fromang_etal12}.

For the properties of the disk outflow, we find
\begin{itemize}
\item The slow magnetosonic point and the Alfv\'en point are always contained
in our simulation box. The location of these critical points shifts towards the
disk midplane as net vertical magnetic flux increases. The launching and
acceleration of the outflow are due to the magnetocentrifugal mechanism, where
surface magnetic field is sufficiently inclined.

\item  There is a robust trend that the outflow mass loss rate $\dot{m}_w$
increases with increasing net vertical field, and tends saturate at
$\beta_0\lesssim10^2$. Although $\dot{m}_w$ can not be well determined from
shearing-box simulations, its exact value in real systems is likely to be much
smaller than the measured from local simulations and may depend on the
aspect ratio $H/R$.

\item The outflow from the MRI simulations is unlikely to be {\it directly} connected
to a global disk wind for geometric reasons. For $\beta_0\gtrsim10^3$, the large
scale radial and azimuthal field lines are constantly flipped due to the dynamo
activities without a permanent bending direction. For $\beta_0\lesssim10^3$, the
large scale magnetic fields do not change sign across the midplane, hence the
field lines at the two sides of the disk bend to opposite directions, inconsistent
with a global wind geometry.

\item Angular momentum transport by the disk outflow is likely to play a minor
role compared with that by the MRI turbulence for relatively weak net vertical
magnetic flux. With stronger net vertical flux ($\beta_0\lesssim10^3$), the
symmetry of the flow (as the dynamo is suppressed) makes net vertical angular
momentum transport be zero, but gives oppositely directed inflow/outflow with
substantial mass flux.
\end{itemize}

It has been reported that $\dot{m}_w$ from shearing-box MRI simulations
decreases with increasing height of the simulation domain, which is related
to the intrinsic limitations of the shearing-box framework. We propose that a
careful study of the height dependence can help determining the true value
of $\dot{m}_w$ in real systems.
Our results leave an open question on the fate of the outflow, which can only
be reliably explored using global simulations and will be discussed further in
Section \ref{ssec:winddiscussion}.

\subsection[]{Implications for Global Disk Evolution}\label{ssec:global}

We note that shearing-box simulations with zero net vertical flux tend to
give $\alpha\sim0.01-0.02$ with confirmed numerical convergence
\citep{MillerStone00,Hirose_etal06,Shi_etal10,Davis_etal10,Simon_etal12a}.
The vast dominance of simulations of this type has, to some extent, implicitly
created a misconception that the $\alpha$ value resulting from the MRI
turbulence if of the order $0.01$. On the observational side, the value of
$\alpha$ can be estimated from the viscous timescale of transient systems
such as dwarf novae, X-ray transient and FU Ori bursts, where it was found
to be of the order $0.1$ or larger \citep{Smak99,Dubus_etal01,Zhu_etal07}.
Such apparent discrepancy leads \citet{King_etal07} to question about the
efficiency of the MRI in driving disk accretion and evolution. The dependence
of $\alpha$ on net vertical magnetic flux has already been extensively
discussed in unstratified shearing-box simulations
\citep{HGB95,Sano_etal04,Pessah_etal07,LongarettiLesur10}. Our
simulations further confirm and quantify this trend using more realistic
simulations with vertical stratification, and demonstrate that the above
discrepancy can be naturally resolved if accretion disks are threaded by
some large scale poloidal magnetic fields.

The strong dependence of $\alpha$ on $\beta_0$ immediately indicates
that the evolution of accretion disks strongly depends on the distribution
of poloidal magnetic flux through the disk. On the other hand, because of
flux freezing in ideal MHD, the distribution of magnetic flux is intimately
connected to disk evolution (as well as turbulent diffusion and reconnection
due to the MRI). Therefore, the mutual dependence of disk evolution and
magnetic flux evolution makes the problem of disk accretion intrinsically
non-local, hence a full understanding of accretion disk structure and
evolution requires global simulations.

Most global simulations to date adopt a magnetic field geometry that is
analogous to zero net-flux shearing-box: the simulations are initialized
either with single/multiple poloidal magnetic loops for a thick disk torus
\citep{Hawley00,DeVilliers_etal03,Penna_etal10}, or with pure toroidal
magnetic field in a thin power-law disk
\citep{FromangNelson06,Beckwith_etal11,Flock_etal11}. These simulations
generally show values of $\alpha$ that are slightly higher than $0.01$
($\sim0.025$). In these simulations, the MRI generates net vertical
magnetic flux in the local patches of the disk connected by coronal loops
and the local stress correlates with the net vertical flux
\citep{ToutPringle92,Sorathia_etal10}.
Combined with our study, the local net vertical flux in these global simulations
is sufficient to account for the slightly higher total stress $\alpha$ than that in
zero net flux shearing-box simulations. However, the distribution of local net
vertical magnetic flux falls off quickly with stronger net flux, hence the total
value of $\alpha$ can not be enhanced substantially.

Initial magnetic field geometry plays an important role in global simulations
of the MRI, although the situation with large scale poloidal fields threading
the disk have rarely been explored. Semi-analytical treatment reveals that
the global distribution of large-scale magnetic flux in accretion disks depends
largely on the ratio of turbulent viscosity and resistivity \citep{Lubow_etal94a},
as well as the vertical structure of the disk \citep{GuiletOgilvie12}. Together
with these works, our simulations provide strong motivation for exploring the
consequence of large scale poloidal field in global simulations. Such
simulations will not only expand our knowledge the MRI, but will also address
the fate of the outflow, which we discuss in the next section.


\subsection[]{Disk Wind Launching in Global Simulations?}\label{ssec:winddiscussion}

It has been recognized that strong net vertical field that approaches
equipartition strength at the disk midplane is necessary for launching a
steady-state wind \citep{WardleKoenigl93,FerreiraPelletier95,OgilvieLivio01}.
The requirement may be relaxed to lower magnetizations \citep{Murphy_etal10},
while for magnetization that is too low the outflow is either suppressed or
highly unsteady \citep{Tzeferacos_etal09}.
Numerous global simulations have studied launching of disk wind in the context
of protostellar disks (e.g., \citealp{Kato_etal02,CasseKeppens02,Zanni_etal07})).
However, as noted in Section \ref{sec:intro}, the use of artificial diffusion in these
simulations do not properly reflect the disk microphysics.
For strong net vertical field that would suppress the MRI, the artificial
diffusion has unknown origin (and the non-ideal MHD effects can not be properly
treated as artificial diffusion, \citealp{BaiStone12b}). For weaker net vertical field,
the MRI is the presumed source of artificial diffusion. However, our evidence
against direct wind launching all originate from the microphysical effects of the
MRI (such as the MRI dynamo) that can not be captured in these simulations.

Another family of global simulations focus on non-radiative accretion disks
around black holes (BH). These simulations either use psudo-Newtonian
gravitational potential for the BH or adopt a general-relativistic prescription.
Despite the non-Keplerian nature in the vicinity of the BHs, the physics of the
MRI that is not too close to the innermost stable orbit is similar. Most of these
simulations adopt similar poloidal loop field geometry and focus on jet
launching which deals with the innermost part of the accretion flow, a few of
them do have included large scale poloidal fields. In particular,
\citet{Beckwith_etal09} initialized the simulations with pure vertical magnetic
fields with $\beta_0\sim10^2$. However, upon saturation the launching of a
Blandford-Payne type wind was not observed, which was attributed to a
global ``coronal mechanism" that governs the magnetic flux evolution. Also,
their simulations have relatively small radial domain and very thick disk.
Significant outflow has been reported in two-dimensional MHD simulations
\citep{StonePringle01,Yuan_etal12b}, where MRI is a transient phenomenon.
More recently, \citet{Tchekhovskoy_etal11}, \citet{McKinney_etal12} and
\citet{Narayan_etal12} studied the accretion flow with large scale initial poloidal
field in the inner disk with $\beta_0\sim10^2$ which leads to a ``magnetically
arrested" or ``magnetically chocked" accretion flow due to the accumulation of
magnetic flux in the vicinity of the BH. The MRI in the main disk was properly
resolved in these simulations, however, they do not report any significant disk
outflow in the inner region of the disks, and found that the jet power is directly
correlated with BH spin.

Together, it appears that Blandford-Payne type disk wind has not been
unambiguously identified in global simulations with properly resolved MRI
turbulence. Future studies should focus on the largely unexplored area of
3D global simulations for thin accretion disks in the presence of large-scale
poloidal magnetic field.
Large simulation domain in both radius and polar angle, fine resolution in the disk,
and careful treatment of boundary conditions are required (T. Suzuki, private
communication), and such simulations are essential to ascertain the nature of the
outflow, and to reveal the underlying physics governing the evolution of accretion
disks.

\acknowledgments

We thank Sebastian Fromang and Geoffroy Lesur for sharing with us their
submitted manuscripts, and the referee, Geoffroy Lesur for helpful
comments that improve the clarity of this paper. We also acknowledge Adam
Burrows, Sebastian Fromang,  Shu-ichiro Inutsuka, Anders Johansen,
Geoffroy Lesur, Ramesh Narayan, Gordon Ogilvie, Dimitrios Psaltis,
Aleksander Sadowski, Takeru Suzuki, Taku Takeuchi, Alexander Tchekhovskoy
and Feng Yuan for useful discussions. XNB acknowledges support for program
number HST-HF-51301.01-A provided by NASA through a Hubble Fellowship grant
from the Space Telescope Science Institute, which is operated by the Association
of Universities for Research in Astronomy, Incorporated, under NASA contract
NAS5-26555. JMS acknowledges support from the National Science Foundation
through grant AST-0908269. This work used computational facilities provided by
PICSciE at Princeton University.


\bibliographystyle{apj}


\label{lastpage}
\end{document}